\documentclass[traditabstract]{aa} 
\usepackage{graphicx}
\usepackage{subfig}
\usepackage{txfonts}
\usepackage{natbib}
\newcommand{\Hi}{\textup{H\,{\mdseries\textsc{i}}}}
\newcommand{\HI}{\textup{H\,{\mdseries\textsc{i}}}~}

\newcommand{\Msun}{{$M_{\odot}$}}
\newcommand{\Lsun}{{$L_{\odot}$}}

\def\kms{{km~s$^{-1}$}}

\def\degr{\hbox{$^\circ$}}

\def\arcsec{\hbox{$^{\prime\prime}$}}
\def\fdegr{\hbox{$.\!\!^\circ$}}

\def\farcsec{\hbox{$.\!\!^{\prime\prime}$}}
\begin{document}
   \title{Cold gas in massive early-type galaxies: The case of NGC 1167}

   \subtitle{}

   \author{C. Struve\thanks{e-mail: struve@astron.nl}
          \inst{1,2},
          T. Oosterloo\inst{1,2},
          R. Sancisi\inst{2,3},
          R. Morganti\inst{1,2}
          \and
          B.H.C. Emonts\inst{4}
          }


   \institute{Netherlands Institute for Radio Astronomy,
              Postbus 2, 7990~AA, Dwingeloo, The Netherlands
         \and
             Kapteyn Institute, University of Groningen,
             Landleven 12, 9747~AD, Groningen, The Netherlands
         \and
             INAF -- Observatorio Astronomico di Bologna,
             via Ranzani 1, 40127 Bologna, Italy
	 \and
	     Australia Telescope National Facility, CSIRO Astronomy and Space Science,
             PO Box 76, Epping, NSW, 1710, Australia
             }

   \date{Received 30 June 2010; accepted 31 August 2010}

  \abstract {We present a study of the morphology and kinematics of the neutral hydrogen in the gas-rich ($M_{\rm{HI}}=1.5\times 10^{10}$~\Msun ), massive early-type galaxy NGC~1167, which was observed with the Westerbork Synthesis Radio Telescope (WSRT). The \HI is located in a 160~kpc disk ($\approx 3\times D_{\rm{25}}$) and has low surface density ($\le 2$~\Msun~pc$^{-2}$). The disk shows regular rotation for $r<65$~kpc but several signs of recent and ongoing interaction and merging with fairly massive companions are observed. No population of cold gas clouds is observed -- in contrast to what is found in some spiral galaxies. This suggests that currently the main mechanism bringing in cold gas to the disk is the accretion of fairly massive satellite galaxies, rather than the accretion of a large number of small gas clumps.
NGC~1167 is located in a (gas-) rich environment: we detect eight companions with a total \HI mass of $\sim$$6\times 10^9$~\Msun\ within a projected distance of 350~kpc. Deep optical images show a disrupted satellite at the northern edge of the \HI disk. The observed rotation curve shows a prominent bump of about 50~\kms\ (in the plane of the disk) at $r\approx 1.3\times R_{\rm{25}}$. This feature in the rotation curve occurs at the radius where the \HI surface density drops significantly and may be due to large-scale streaming motions in the disk. We suspect that both the streaming motions and the \HI density distribution are the result of the interaction/accretion with the disrupted satellite. Like in other galaxies with wiggles and bumps in the rotation curve, \HI scaling describes the observed rotation curve best. We suggest that interactions create streaming motions and features in the \HI density distribution and that this is the reason for the success of \HI scaling in fitting such rotation curves.}


   \authorrunning{C. Struve et al.}

   \maketitle

\section{Introduction}

Observations show that early-type galaxies (ETGs) continue to form their structures over cosmic times down to the present day. Evidence comes from studies of ionised gas \citep[e.g.][]{sadler85,sarzi06}, fine structure in optical morphology \citep[e.g.][]{malin83,schweizer92} and from stellar-population investigations \citep[e.g.][]{trager00,tadhunter05}. Numerous studies of the properties of \HI in and around ETGs \citep[see, e.g.,][]{knapp85, roberts94, grossi09} are strongly supporting the picture of an on-going assembly of ETGs. 
About 50\% of the field ETGs have detectable amounts of \Hi, sometimes forming large, regular rotating disk structures \citep[e.g.][]{morganti06, oosterloo07b}.

In order to be able to evolve, ETGs need to keep on accreting cold gas. For instance, cold gas is required to explain the stellar kinematical properties observed in ETGs \citep{emsellem07}. \citet{serra10} find evidence that up to 5\% of the stellar mass in ETGs is contained in young stars that formed during the past $\sim$1~Gyr. Moreover, \citet{serra10} argue that a large fraction of ETGs have accreted about 10\% of their baryonic mass as cold gas over the past few Gyr.

In addition, theoretical work suggests that cold gas is an essential ingredient for the evolution of ETGs. While dissipationless (gas-free or ``dry'') mergers can explain boxy, slowly rotating massive galaxies \citep[e.g.][]{faber97,kormendy09}, a comparison of N-body simulations with observations suggests that present-day ETGs can not have assembled more than 50\% of their mass via dry merging \citep{nipoti09}. \citet{jesseit07} and \citet{hopkins09} have shown that simulations of dry mergers are not able to reproduce some observed quantities such as counter-rotating cores, velocity dispersion features and structural properties of ETGs, whereas the inclusion of a gas component in the simulations (10\% of the baryonic mass) helps to reproduce the main observed features. Moreover, dissipative mergers and satellite accretion events are needed to explain, e.g., the dynamical structure of disk-like ETGs \citep{naab06}.

The accretion/merging of fairly massive companion galaxies is one possible process to provide the required gas masses to ETGs, but other mechanisms can be important as well. In particular the slow but long-lasting infall of cold gas from the IGM is predicted as a possible mechanism which can provide large gas masses over cosmic times \citep[e.g.][]{binney77, keres05}. In the framework of hierarchical galaxy formation, new cold gas is added to the halo or deposited in the outer parts of galaxies where it forms a reservoir to fuel the inner parts of galaxies. However, the exact mechanism, how the gas cools (or remains cool), condenses and penetrates to the inner disk without getting shock heated is currently still under debate \citep[e.g.][]{peek08, dekel08, kawata08, brooks09, keres09}.

Whether the accretion/merging of a few fairly massive satellites is the main disk building process or whether the continuous infall of many small gas clumps plays a dominant role for the continuing assembly of ETGs is not clear. Given the importance of cold gas for the evolution of ETGs, motivated by observations and theoretical work, it is essential to investigate the origin of cold gas in ETGs.

The question of how galaxies accrete gas is not only important for ETGs, but also for spiral galaxies and hence for galaxy evolution as a whole. Signs of gas accretion are found in a large number of galaxies \citep[for instance: gas-rich minor mergers, lobsided disks, gas filaments, extra-planar \Hi; for a review, see][]{sancisi08}, showing the continuing assembly of galaxies. For spiral galaxies the continuous supply of cold gas is important because gas consumption times are usually short in the inner regions \citep[typically $\le1-2$~Gyr; see, e.g.,][]{bigiel08}. Hence, these galaxies can only maintain their observed star formation rates if they manage to keep on acquiring gas, either through merging or via the accretion of small gas clumps from the IGM. An interesting aspect in that respect is whether ETGs accrete gas in the same way, or if other mechanisms might suppress the accretion of cold gas onto the disk, which may then explain the absence of current star formation.

Neutral hydrogen studies of gas-rich, massive ETGs provide a powerful tool to investigate where the cold gas in these systems comes from. Because the cold gas column densities must be low (since large-scale star formation does not occur), the \HI in \Hi-rich ETGs is often distributed in large structures \citep[see, e.g.,][]{morganti06, oosterloo07b}. At large radii, dynamical timescales are large, allowing to investigate the evolution of the galaxy over the past few Gyr. Since ETGs continue to form their structures down to the present day, the role of the \HI in this formation process can be investigated over Gyr timescales.

A second reason to observe \Hi-rich ETGs is that it allows to study in detail the accretion of small gas clumps. In recent years, neutral hydrogen gas has been detected in the halo of spirals \citep[for a review, see][]{sancisi08} and the existence of \HI in the halo could be partly explained by the accretion of small cold gas clumps. However, an alternative mechanism for bringing \HI in the halo is star formation (supernova explosions and stellar winds blow gas into the halo) and it is not easy in spiral galaxies to determine which of the two mechanism is the most important one.

To study the accretion of small gas clouds, it is thus useful to observe galaxies with no or little star formation, such as early-type and low surface brightness galaxies that have no/few supernovae explosions. Only very few such galaxies have been studied so far. \citet{matthews03} detected \HI in the halo of the LSB galaxy UGC~7321 but concluded that even in this galaxy the energy provided by supernovae explosions might be sufficient to explain the gas in the halo.

%

In this paper we investigate how the cold gas in the early-type (SA0) galaxy NGC~1167  (UGC~2487) is accreted in order to understand where the \HI comes from. We present deep \HI observations performed with the Westerbork Synthesis Radio Telescope (WSRT). NGC~1167 is a giant disk galaxy ($M_{\rm{B}} = -21.7$~mag, $D_{25}=56$~kpc), seen at low inclination, known to host a relatively powerful ($10^{24}$~W~Hz$^{-1}$ at 1.4~GHz) radio source (B2~0258$+$35). What makes this galaxy interesting for the study of the origin of the cold gas is that previous \HI observations had shown a regular rotating $160$~kpc disk with a total \HI mass $>10^{10}$~\Msun\ \citep{noordermeer05,emonts07}. The \HI surface density is low ($\le 2$~\Msun~pc$^{-2}$) and the \HI disk extends beyond the optical disk out to ten $R$-band disk scale lengths. Deep optical observations show a faint, tightly wound spiral structure at $r<30$~kpc \citep{emonts10}. No stellar bar structure is visible in the data. H$\alpha$ emission-line imaging revealed no emission (${\rm{log}}~L < 37.6$), except for close to the AGN ($r<5$~kpc), and hence no sign of ongoing star formation \citep{pogge93} -- in agreement with H$\alpha$ long-slit observations obtained by \citet{emonts06}. X-ray observations at 2 -- 10~keV \citep{akylas09} show a X-ray halo which extends beyond $D_{25}$ (J. Rasmussen, private communication). In addition to the investigation of the origin of the cold gas, the very extended \HI disk allows to derive a rotation curve out to large radii and to investigate the mass distribution out to large distances from the centre. Some optical and radio parameters are given in Table~\ref{general.properties}.

The organisation of this paper is as follows: in Sect.~2 we describe the observations, while Sect.~3 presents the observational results and in Sect.~4 we search for a possible \HI halo component. In Sect.~5 we derive a rotation curve to investigate the disk dynamics. A discussion is given in Sect.~6 and a summary in Sect.~7. The redshift of NGC~1167 ($z=0.0165$) translates to a distance of $\rm{D}=70$~Mpc assuming $H_{\rm{0}}=71$~\kms ~and consequently 3\arcsec ~correspond to about 1~kpc.

%
\begin{table}
\caption{Optical and radio parameters for NGC~1167.}
\label{general.properties}
\centering
\begin{tabular}{l c c}
\hline\hline
Parameter & Value & Reference\\
\hline
Morphological type & SA0 & 1\\
Center $\alpha$ (J2000.0) & 03 01 42.4 & 1\\
Center $\delta$ (J2000.0) & 35 12 20.7 & 1\\
Distance (Mpc) & 70 & 1\\
$L_B$ (\Lsun ) & $7.4\times 10^{10}$ & 2\\
Systemic velocity (\kms ) & $4951\pm 3$ & 3\\
$D_{25}$ (kpc) & 49 & 1\\
$D_{\rm{HI}}$ (kpc) & 160 & 3\\
Total \HI mass (\Msun ) & $1.5\times 10^{10}$ & 3\\
\HI inclination (deg) & 38 & 3\\
Mean P.A. (no warp) (deg) & 75 & 3\\
Total mass (\Msun) & $>1.6\times 10^{12}$ & 3\\
\hline
\end{tabular}
\caption*{Note. -- Units of right ascension are hours, minutes, and seconds and units of declination are degrees, arcminutes, and arcseconds.
References. -- (1) NED; (2) based on LEDA; (3) this work.}
\end{table}
%
%
\begin{table}
\caption{Summary of observations.}
\label{obs}
\centering
\begin{tabular}{c l}
\hline\hline
Total observing time & $138$~h\\
Total on-source integration time  & $\sim 107$~h\\
Observing period & Oct. 2006 to\\
 & Nov. 2008\\
Bandwidth & 20~MHz\\
Number of spectral channels & 1024\\
Channel separation & 8.5~\kms\\
Velocity resolution (after Hanning) & 17.0~\kms\\
Bandpass/flux calibrator & 3C48 and 3C147\\
\hline
Weighting scheme & robust 0.4\\
Resolution & 28\farcsec 9~x~16\farcsec 8\\
Beam position angle & 1\fdegr 5\\
rms noise (mJy~beam$^{-1}$) in the cubes & 0.125\\
rms noise per channel ($10^{18}$atoms~cm$^{-2}$) & 4.8\\
rms noise per channel (\Msun ~pc~$^{-2}$) & 0.039\\
Detection limit ($3\sigma$ over 3 channels) ($10^6$\Msun ) & 5.6\\
\hline
\end{tabular}
\end{table}
%
%
\begin{figure*}
\centering
\includegraphics[width=0.75\textwidth]{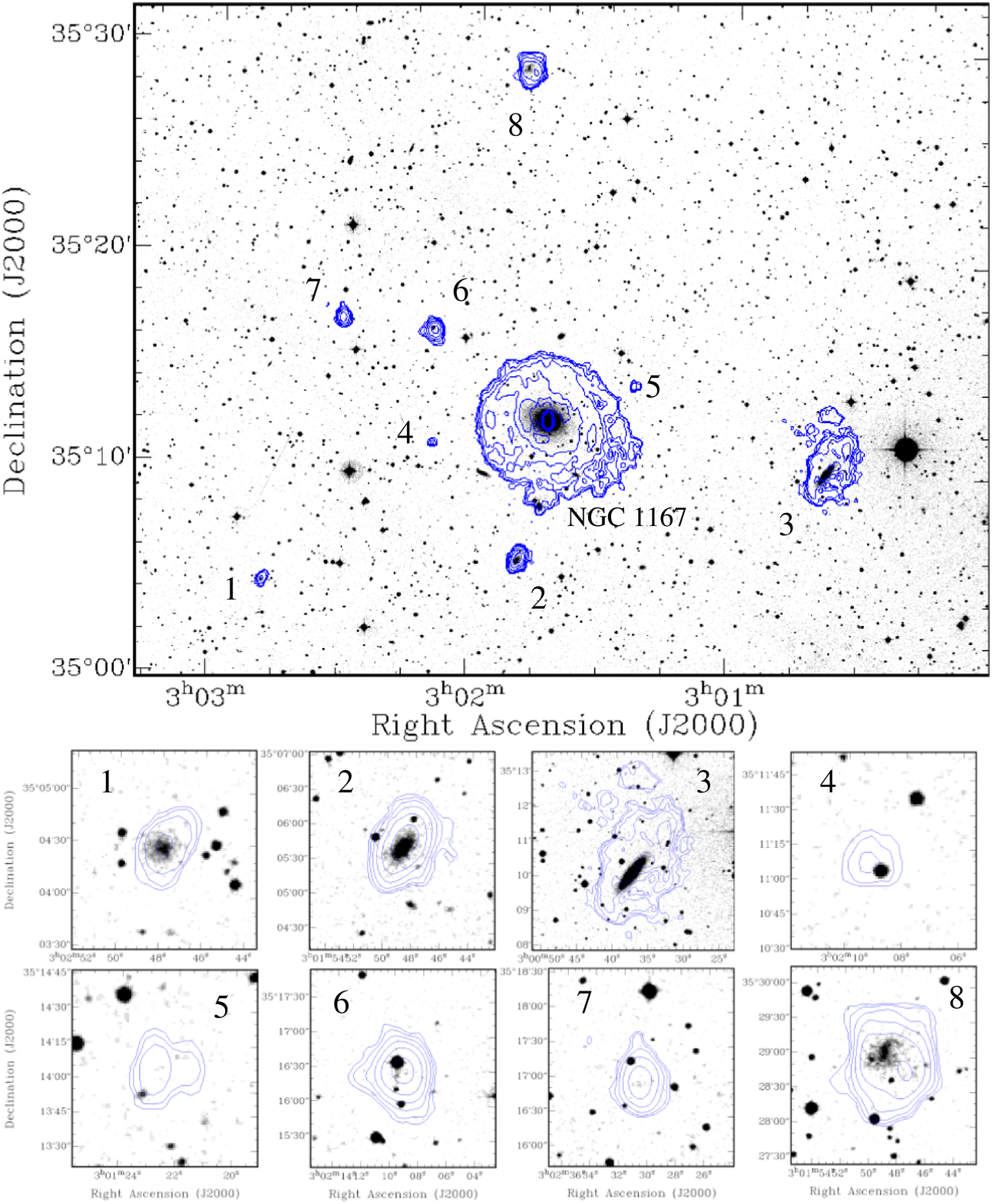}\\
\caption{DSS2 image showing the \HI group environment of NGC~1167. Numbers correspond to numbers in Table~\ref{neighbours}. \HI contour levels are: 1, 2, 4, 8, 16 and $32\times 10^{19}$~atoms~cm$^{-2}$.}
\label{mom0}
\end{figure*}
%


\section{Observations and data reduction}

A ten-hour WSRT service observation in 2006 was followed-up by eleven 12-h observing runs in 2008 (see Tab.~\ref{obs}). The net on-source integration time is about 107~h. The strong radio continuum source in the centre (1.8~Jy) required non-standard calibration. In order to guarantee a stable bandpass, calibration observations of 15 minutes were performed every two hours. 3C48 and 3C147 were used as calibrators. An almost uniformly filled uv-plane could be achieved for the combined data by interleaving the hour angle of the calibration observations between individual observations.

%
\begin{table*}
\caption{Properties of \HI detections.}
\label{neighbours}
\centering
\begin{tabular}{c r c c c r r r r r r r r}
\hline\hline
No. & Identification & RA & Dec & $v_{\rm{sys}}$ & $\Delta v$ & $\Delta v_{\rm{off}}$ & $D$ & $D_{\rm{off}}$ & $D_{\rm{off}}$  & $F_{\rm{o}}$  & $F_{\rm{c}}$ & $M_{\rm{HI}}$ \\
(1) & (2) & (3) & (4) & (5) & (6) & (7) & (8) & (9) & (10) & (11) & (12) & (13)\\
\hline
0 & NGC~1167 & 03 01 42.4 & 35 12 20.7 & 4951 & 520 & - & 69.7 & - & - & 457.65 & 461.33 & 15.324 \\
\hline
1 & J0302478+3504284$^1$ & 03 02 47.8 & 35 04 28.4 & 4864 & 103 & -87 & 68.5 & 15.5 & 310.4 & 1.62 & 2.54 & 0.081\\
2 & 2MASX$^2$ & 03 01 48.6 & 35 05 42.4 & 4903 & 213 & -48 & 69.1 & 6.7  & 134.8 & 16.29 & 17.70 & 0.578\\
3 & UGC 2465 & 03 00 37.4 & 35 10 08.7 & 5065 & 529 & 114 & 71.3 & 13.4 & 267.6 & 57.23 & 80.90 & 2.812\\
4 & J0302091+3511120$^3$ & 03 02 09.1 & 35 11 12.0 & 4907 & 34 & -44 & 69.1 & 5.6 & 111.6 & 0.39 & 0.41 & 0.013\\
5 & J0301222+3514055$^4$ & 03 01 22.2 & 35 14 05.5 & 5057 & 42 & 106 & 71.2 & 4.5 & 89.4 & 0.78 & 0.81 & 0.028\\
6 & J0302095+3516308$^3$ & 03 02 09.5 & 35 16 30.8 & 4903 & 77 & -48 & 69.1 & 6.9 & 138.0 & 13.20 & 14.39 & 0.470\\
7 & J0302307+3516581$^4$ & 03 02 30.7 & 35 16 58.1 & 5245 & 128 & 194 & 73.9 & 10.9 & 217.9 & 6.07 & 7.56 & 0.282\\
8 & J0301491+3529012$^1$ & 03 01 49.1 & 35 29 01.2 & 4954 & 94 & 3 & 69.8 & 16.7 & 334.0 & 25.59 & 43.10 & 1.436\\
\hline
\end{tabular}
\caption*{Note. -- (1) Number in Fig.~\ref{mom0}. (2) Name (NED)/Identification. (3) Centre: Right ascension. (4) Centre: Declination. (5) Systemic velocity [\kms]. (6) Line width above $3\sigma$. (7) Velocity off-set from NGC~1167. (8) Distance from Earth assuming Hubble flow ($H_0=71$) [Mpc]. (9) Off-set from NGC~1167 [arcmin]. (10) Off-set from NGC~1167 [kpc]. (11) Integrated flux [Jy~\kms]. (12) Primary beam corrected Flux [Jy~\kms]. (13) \HI mass [$10^9$\Msun]. $^1$ counterpart on DSS image; $^2$ full name is: 2MASX~J03014864+3505423; $^3$ star in front; $^4$ no counterpart on DSS image.}
\end{table*}
%
%
\begin{figure}
\centering
\includegraphics[width=0.33\textwidth, angle=270]{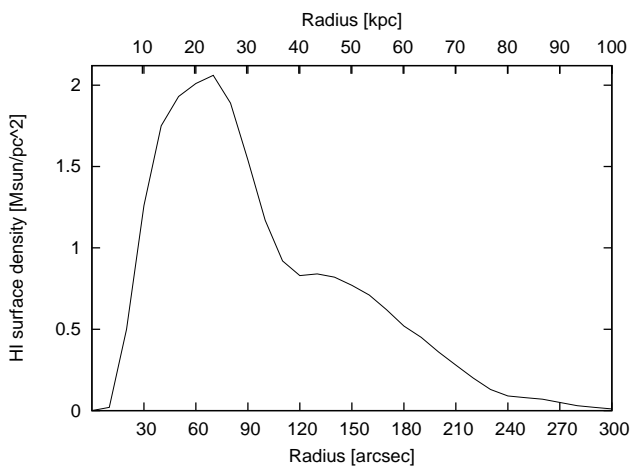}
\caption{Radial \HI surface density profile.}
\label{density}
\end{figure}
%

The observing band was centred on the redshift of NGC~1167. The total bandwidth was 20~MHz, corresponding to a velocity range of $\sim$3700~\kms , and was covered with 1024 channels (dual polarisation). In order to increase the signal-to-noise ratio, two channels were binned and subsequently Hanning smoothed resulting in a spectral resolution of 17~\kms\ in the output data cube. The data reduction was performed using the MIRIAD package \citep{sault95}.

After flagging, bandpass and phase calibration were performed. In addition, on every individual data set self-calibration of the continuum emission was done and the results of this also applied to the line data. The continuum model was subtracted from the line data and the residual continuum emission was removed by making a fit to the line-free channels of each visibility record. The data cube was cleaned using the Clark algorithm. 
For the extended emission a cube with robust weighting of 0.4 gives the best compromise between spatial resolution and sensitivity, which is used for the analysis of the \HI data.


%
\begin{figure}
\centering
\includegraphics[width=0.33\textwidth, angle=270]{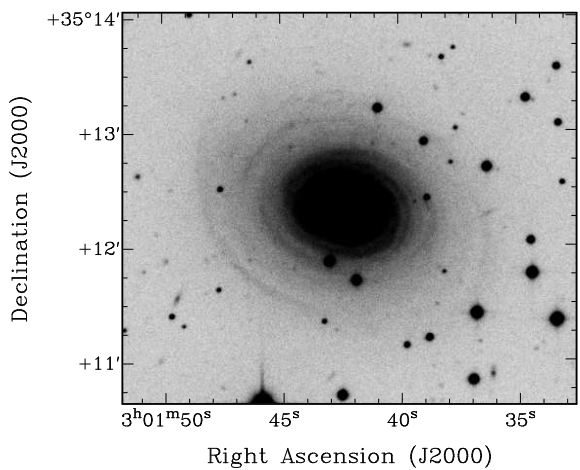}
\caption{ $B$-band image from \citet{emonts10} showing the faint spiral structure of the stellar disk.}
\label{faint.spiral}
\end{figure}
%

\section{Results}
\subsection{NGC~1167}

The high sensitivity of our observations has allowed to detect new faint \HI structures compared to earlier observations. Figure~\ref{mom0} shows the total \HI image of NGC~1167 and its neighbours. This image was obtained by smoothing the data cube to 35\arcsec ~resolution and using this smoothed cube to create a mask (blanking all signal below $+3\sigma_{\rm{rms}}$ in the smoothed cube) which was applied to the original cube. The \HI extends over 160~kpc in diameter ($\approx 3\times D_{25}$), with a total \HI mass detected in emission of $1.5\times 10^{10}$~\Msun ~\citep[which is roughly three times the \HI mass of the Milky Way;][]{henderson82}. Channel maps are shown in Fig.~\ref{channelmaps} and the radial \HI density profile in Figure~\ref{density}. The highest \HI surface density is found at the radius of the faint, tightly wound spiral structure found in optical images (Fig.~\ref{faint.spiral}; see also Emonts et al. 2010). Absorption is detected against the strong central continuum, which is unresolved at our resolution. Higher resolution observations show that the absorption is detected against a kpc scale structure. This will be discussed in a future paper (Struve et al. in prep.). The apparent central drop in surface density (Fig.~\ref{density}) is affected by the absorption. Despite the large \HI mass, the surface density is too low for star formation to occur globally.

\begin{figure*}
\centering
\includegraphics[width=0.38\textwidth, angle=270]{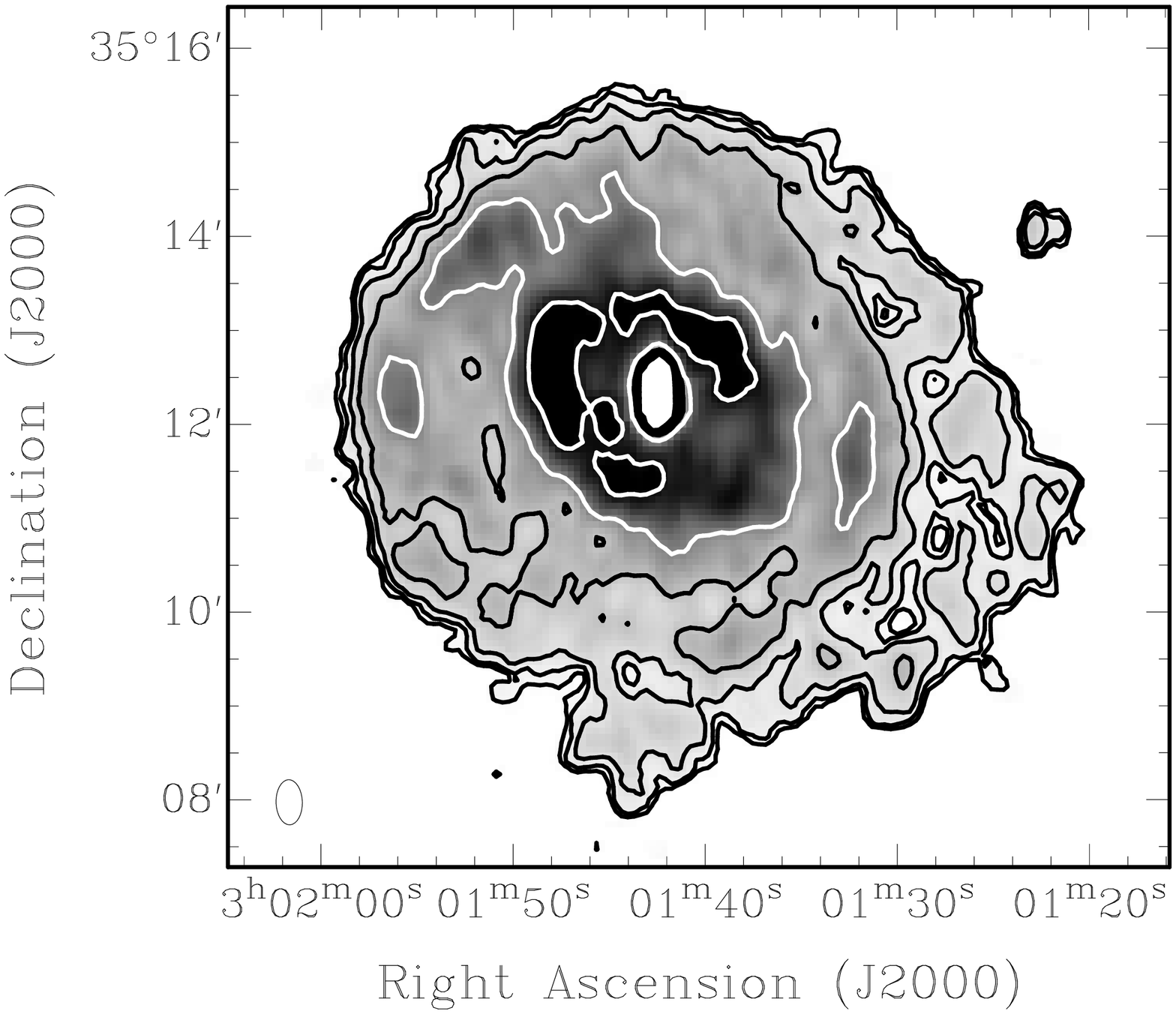}
\includegraphics[width=0.43\textwidth, angle=270]{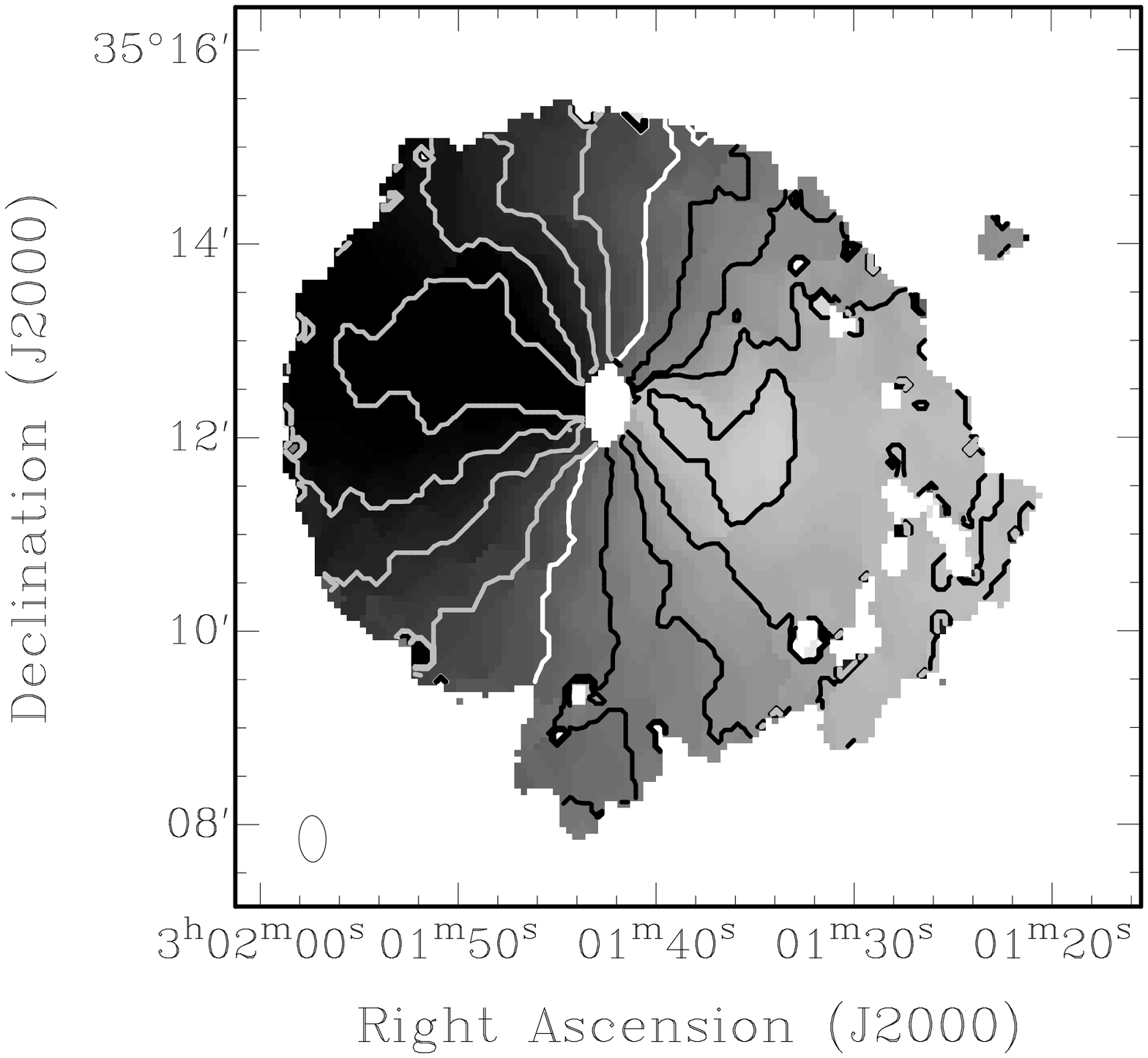}
\caption{Left panel: Total \HI intensity map of NGC~1167. Contour levels are: 1, 2, 4, 8, 16 and $32\times 10^{19}$~atoms~cm$^{-2}$. Right panel: Velocity field based on the peak of the profile (see text). The white contour indicates the systemic velocity, 4951~\kms , the black and grey contours give the velocity in increments of 50~\kms . The approaching side is to the East. The beam is indicated in the lower left corners.}
\label{velocityfield}
\end{figure*}

To the southwest of NGC~1167, the \HI extends to larger radii as compared to the rest of the galaxy (see Fig.~\ref{velocityfield}, left panel). The gas in this region does not follow exactly the rotation of the regularly rotating $\sim$65~kpc disk (see also Fig.~\ref{channelmaps}).

%
%
%
\begin{figure*}
\centering
\includegraphics[width=0.44\textwidth, angle=270]{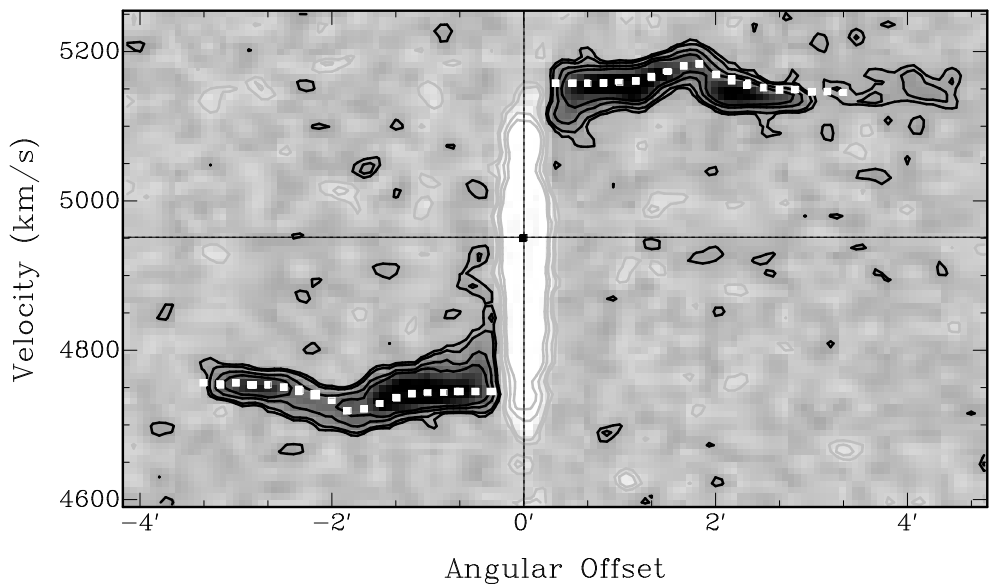}
\caption{Position-velocity slice along the major axis. The white squares give the rotation curve as derived from modelling (see Sect.~5). The bump in the rotation curve is clearly visible. Contour levels: --9, --6, --3, --2 (thin black), 2, 3, 6 and 9$\sigma$ (thick black). The horizontal dashed line indicates the systemic velocity.}
\label{pv.slices}
\end{figure*}
%

\begin{figure*}
\centering
\includegraphics[width=0.37\textwidth, angle=270]{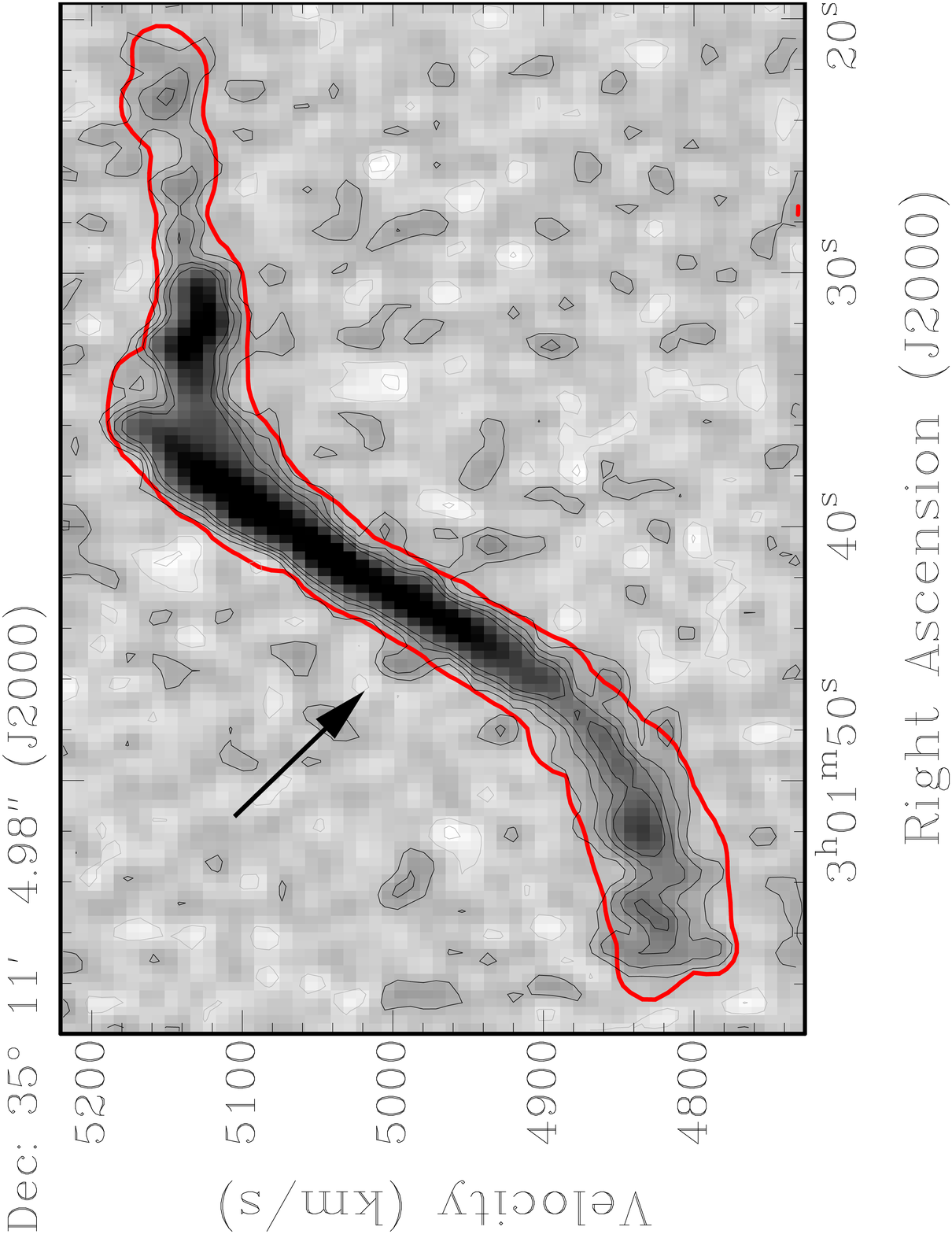}
\includegraphics[width=0.37\textwidth, angle=270]{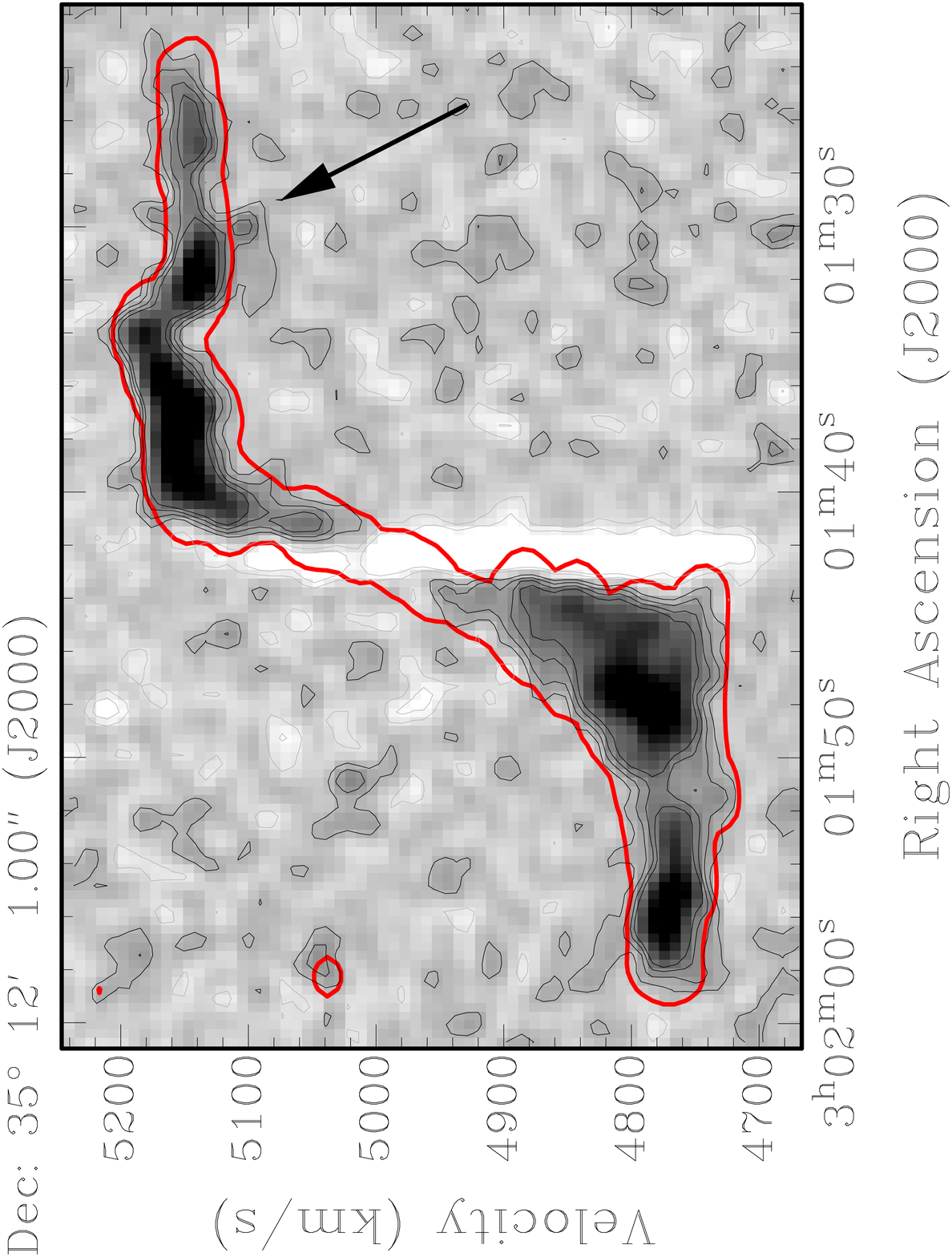}
\caption{Two east-west $pv$-slices of the data cube overplotted with the $1\sigma$ contour level (red contour) of our model. Contour levels of the data: -3, -1.5 (grey), 1.5, 3, 4.5 and $6\sigma$ (black). \HI emission not covered by the model is indicated by arrows.}
\label{pv.kinematicmodel}
\end{figure*}

The derivation of a velocity field is not straightforward. Some of the velocity profiles are broad (see Fig.~\ref{pv.slices} and \ref{pv.kinematicmodel} for a few examples) and close to the centre ($r<100$\arcsec) beamsmearing becomes severe due to the large beam size (9.6~x~5.6~kpc) and the steep velocity gradients. To minimise beamsmearing effects, we use the velocity of the peak of the profile instead of the intensity weighted mean. The resulting velocity field (Fig.~\ref{velocityfield}, right panel) shows that the gas kinematics for $r<65$~kpc is dominated by regular large-scale rotation.

The radial velocities along the major axis and the main rotational properties are shown in Fig.~\ref{pv.slices}. The rotation curve (derived in Sect.~5 and over-plotted on the major-axis slice in Fig.~\ref{pv.slices}) is rising within the central beam and is essentially flat out to the largest radii measured. However, on both the approaching and receding side, at radii between $\sim$26 and $\sim$45~kpc, the rotation curve makes a 25~\kms ~jump up and goes down again by about 37~\kms ~(uncorrected for inclination). The radius of the peak of the jump ($r\approx 36$~kpc) and the radial extent of this jump are similar, but not completely identical for the approaching and receding side. This ``bump'' in velocity occurs just outside the bright optical disk at $r\approx 1.3\times R_{25}$, exactly at the radius where the \HI surface brightness rapidly drops (Fig.~\ref{density}). This is further discussed in Sect.~6.3.

\subsection{The (\Hi ) environment of NGC~1167}

We detect eight sources (Fig.~\ref{mom0}) in the vicinity of NGC~1167 with \HI masses between $\sim 1\times 10^7$~\Msun\ and $3\times 10^9$~\Msun\ within a projected radius of 350~kpc. The total amount of \HI in the environment is $5.7\times 10^9$~\Msun. Two of the detections are identified with catalogued galaxies. Table~\ref{neighbours} summarizes the basic properties of all \HI detections.

\begin{figure}
\centering
\includegraphics[width=0.4\textwidth, angle=270]{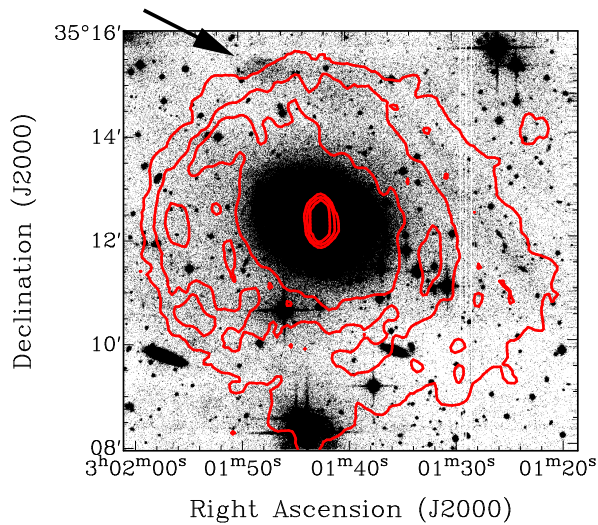}
\caption{$B$-band image from \citet{emonts10} overlaid with \HI density contours. Contour levels: 1, 8 and $16\times 10^{19}$~atoms~cm$^{-2}$. The arrow indicates the location of the disrupted satellite.}
\label{deep.optical}
\end{figure}

The most massive galaxy besides NGC~1167, UGC~2465, is located 268~kpc (projected) west and shows a distorted \HI morphology and kinematics. In addition, deep optical observations \citep[for details, see][]{emonts10} show a disrupted satellite at the northern edge of the \HI disk with a possible tail towards the region of the southwestern \HI extension (Fig.~\ref{deep.optical}, see also Sect.~6.1). Besides these two cases, none of the other \HI detections show hints of an interaction with NGC~1167 in terms of bridges or tails.

Five detected galaxies (No. 1, 2, 6, 7 and 8 in Tab.~\ref{neighbours}) show indications of regular rotation, despite the low spatial resolution. Objects No. 2, 6 and 7 are fairly gas-rich with \HI masses above $10^8$~\Msun . DSS2 images do not show an optical counterpart for No. 6 and 7 (see Fig.~\ref{mom0}). However, the DSS2 images are shallow and optical counterparts with luminosities several times $10^7$~\Lsun\ could still be present. Close to NGC~1167, at distances of 90 to 110~kpc from the centre, we detect two unresolved \HI clouds (No.~4 and 5) with masses just above our detection limit which do not have an optical counterpart in DSS2 either. No additional low-mass \HI clouds were found in a systematic search using the source finder Duchamp \citep{duchamp}.

%
%
%
%

%

\section{The search for an \HI halo}

One of the main reasons to observe NGC~1167 is to investigate the origin of the large amount of \HI in this early-type galaxy. One possibility is the long-term accretion of a large number of small cold gas clumps. Signs for the accretion of such gas clouds have been found in the halos of a few spiral galaxies \citep[for a review, see][]{sancisi08}. Over a Hubble time, galaxies could accrete a significant fraction ($>10^9$~\Msun) of their cold gas in this way. In order to investigate the possibility that NGC~1167 also accretes such gas condensations, we search for an extended halo of \HI clouds. In Sect.~4.1 we explain how to identify gas that is located in the halo and in Sect.~4.2 we show that no \HI clouds are detected in the halo of NGC~1167 and place upper limits to the accretion of small gas clumps.

\subsection{Identifying halo gas}

In low-inclination galaxies, such as NGC~1167, the separation of gas located in the disk and the halo is not straightforward. This is because gas located in the halo lies along the same line-of-sight as disk gas, unlike in edge-on systems. However, the kinematics of the halo gas differs from the disk kinematics such that the velocities of the gas can be used to separate the halo component from disk gas. For instance, it has been found that gaseous halos observed in spirals are lagging with respect to disk rotation \citep{fraternali02, heald07, oosterloo07}. This lagging gas is most clearly seen in a $pv$-diagram along the major axis in the form of wings on the side of the lower rotation velocities, toward systemic. These systematic asymmetries in the \HI velocity profiles have been refered to as {\sl beards} \citep{sancisi01}. Low-inclination galaxies that show prominent beards are e.g. NGC~2403 \citep{fraternali02} and NGC~6946 \citep{boomsma08}.

Position-velocity slices along the major axis of NGC~1167 (Fig.~\ref{pv.slices}) or near the major axis (Fig.~\ref{pv.kinematicmodel}, right panel) show \HI line profiles that are strongly asymmetric with respect to the peak of the line, in particular in the inner part of the disk (similar to NGC~2403 and NGC~6946). Note that the white squares, which mark the rotation curve (which we derive in Sect.~5), follow closely the peaks of the profiles. In some locations the profiles extend (as a beard) 200~\kms\ toward the systemic velocity (Fig.~\ref{pv.kinematicmodel}, right panel).

However, there is a major difference between NGC~1167 and previously studied low-inclination galaxies: NGC~1167 is located relatively far away such that the linear resolution of our observations is coarse (beam is $9.6 \times 5.6$~kpc). As a consequence the velocity gradients within a beam are large. Hence, broad velocity profiles are observed, in particular close to the centre where the velocity gradients are largest. Because of the large rotation amplitude, the velocity gradients would even be larger if NGC~1167 would be observed at the same resolution as the previously studied objects. The convolution of the signal with the observed beam spreads the emission to neighbouring locations such that the broad profiles observed in Fig~\ref{pv.slices} and \ref{pv.kinematicmodel} could be the result of beamsmearing instead of indicating a halo. To identify a possible \HI halo component, it is therefore important to compare the data to a model that takes beamsmearing into account. In addition to the velocity gradients near the centre, NGC~1167 shows evidence for large, local velocity gradients at other locations due to streaming motions in the disk (see Sect.~5). Also these require that beamsmearing is taken into account in modelling the \HI disk of NGC~1167.

The thin disk of NGC~1167 is modeled by placing at every pixel in the cube a Gaussian profile centred at the velocity at that location as derived from the observed velocity field. This cube is then convolved with the beam. Thus, this approach ensures that the local velocity structure and beamsmearing effects are taken into account. The intensity of the profile is taken from the total \HI map at the position. For the Gaussian profile, a velocity dispersion of 9~\kms\ is assumed.

\subsection{An upper limit to the \HI halo}

Comparing the model cube with the data shows that the model describes the overall kinematics of the \HI in NGC~1167, including the broad profiles seen in many locations. This is illustrated in Fig.~\ref{pv.kinematicmodel} where we show two representative $pv$ slices. The fact that our beamsmeared model of the thin \HI disk does explain the broad profiles indicates that these are due to beamsmearing and that we do {\sl not} observe \HI clouds in the halo of NGC~1167, in contrast to what is seen in many later-type spiral galaxies.

Figure~\ref{pv.kinematicmodel} also illustrates that the only features not consistent with our model are at the noise level. However, many other such ``blobs'' are present in the data cube, also at locations and velocities quite remote from NGC~1167. Therefore, we take the masses and column densities of the clouds of the kind indicated in Fig.~\ref{pv.kinematicmodel} as the upper limit for any halo population in NGC~1167.

The upper limit of the peak surface density of a possible \HI halo component is hence $0.05$~\Msun\ pc$^{-2}$, which corresponds to a column density of $6\times 10^{18}$~cm$^{-2}$ and any cloud in the halo of NGC~1167 has an \HI mass of less than $10^7$~\Msun. We note, for comparison, that in some spiral galaxies halos have been observed with average column densities well above $10^{19}$~cm$^{-2}$ while the regions with highest \HI surface density in the halo of, e.g., NGC~2403 and NGC~6946 have column densities above $10^{20}$~cm$^{-2}$ \citep{fraternali02,boomsma08}, which is a factor $\sim$20 above the upper limit for NGC~1167. We can, therefore, rule out that NGC~1167 has an \HI halo with similar (column) densities as compared to those late-type, star-forming galaxies. The mean density contrast ratio between \HI located in the disk and in the halo is $>20$ for NGC~1167. This is significantly larger than for NGC~2403 and NGC~891 (ratio is 10 and 2.3 respectively), but is similar to the one of NGC~6946 (ratio is 22).

\section{Rotation curve}

In this section we further investigate the nature of the velocity bump seen in the kinematics of NGC~1167, as described in Sect.~3.1. In particular, we investigate whether this feature reflects the overall mass distribution of this galaxy, or whether it could be indicative of a recent disturbance suffered by NGC~1167. We derive a rotation curve and show that none of our applied mass models explains the feature in the kinematics satisfactorily. We conclude that large-scale streaming motions must be present in the disk.

\begin{figure}
\centering
\includegraphics[width=0.35\textwidth, angle=270]{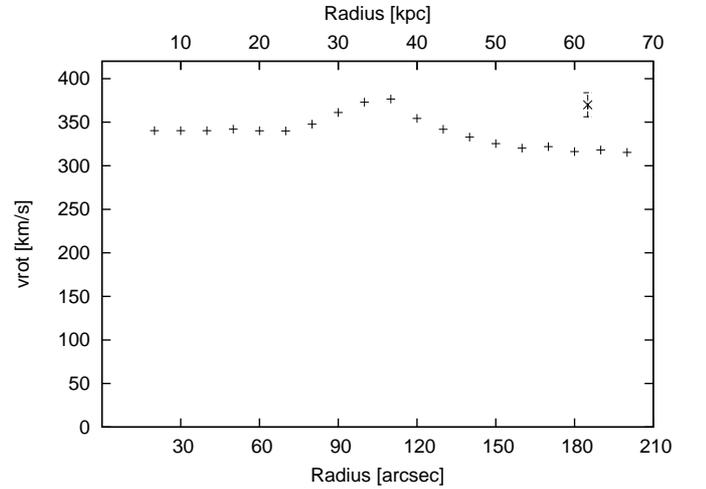}
\caption{Rotation curve. The error bar indicates the uncertainties.}
\label{rotation.curve}
\end{figure}

\begin{figure}
\centering
\includegraphics[width=0.33\textwidth, angle=270]{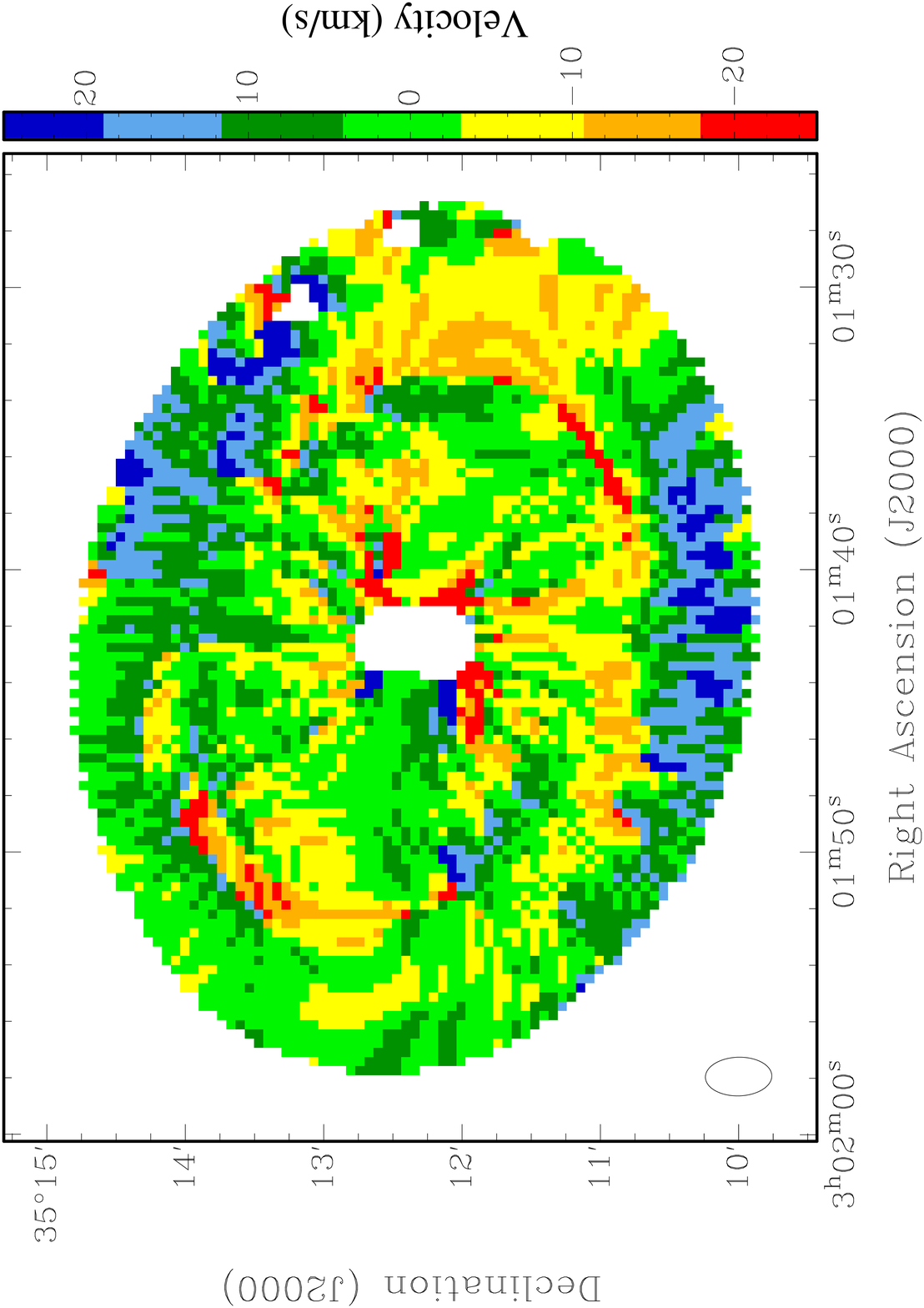}
\includegraphics[width=0.33\textwidth, angle=270]{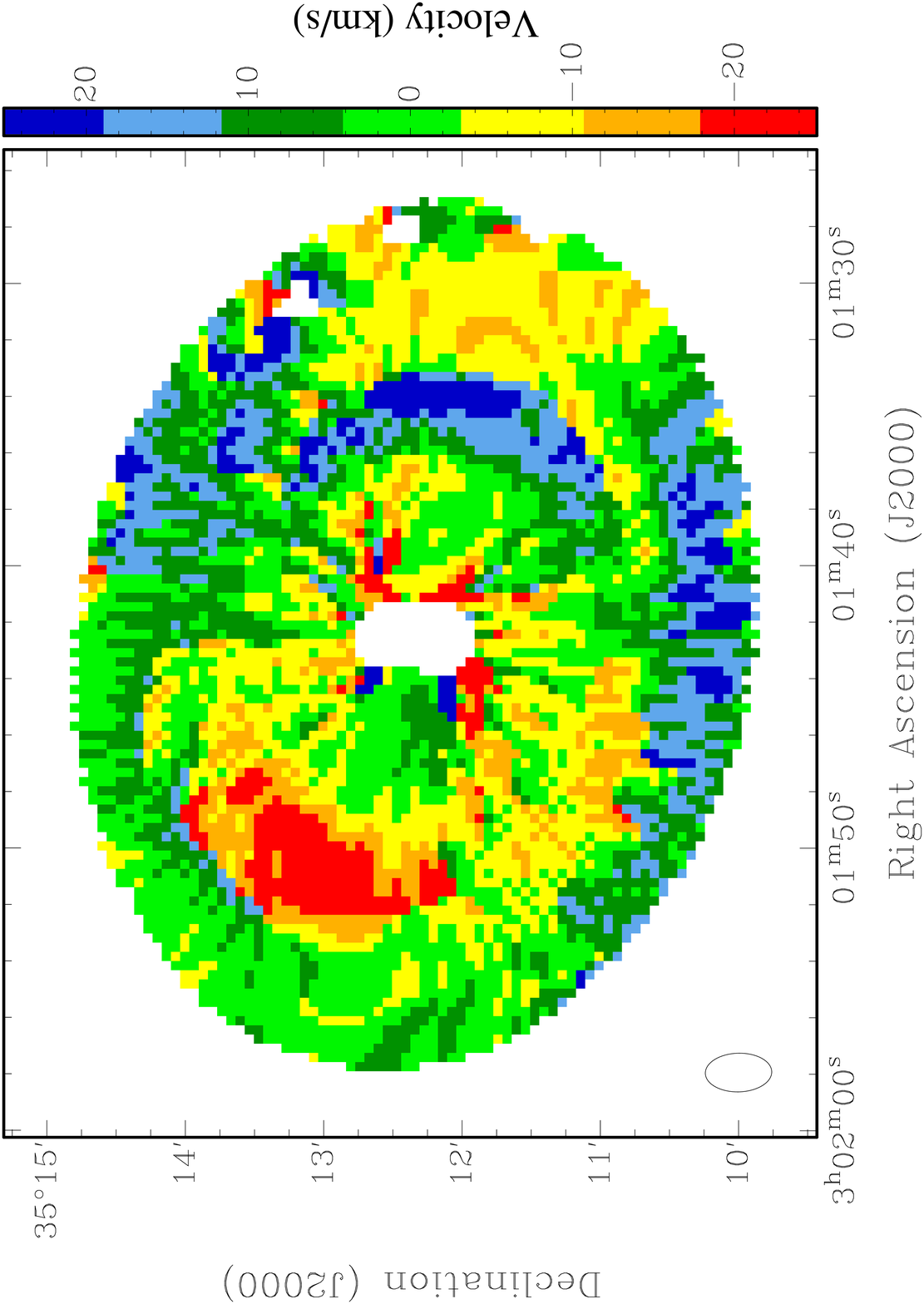}
\caption{Top panel: Residual velocity field based on the derived rotation curve (Fig.~\ref{rotation.curve}). Bottom panel: Residual velocity field assuming a flat rotation curve (see text). The beam is indicated in the lower left corners.}
\label{vel.residuals}
\end{figure}

The rotation curve is derived by fitting a set of tilted rings to the observed velocity field (Fig.~\ref{velocityfield}, right panel) \citep[following][]{begemann87}. The resulting rotation curve is plotted in Fig.~\ref{rotation.curve} and given in Tab.~\ref{RC}. The rotation curve has two sources of uncertainty. Every data point has an individual error of $\sim$13~\kms\ (estimated from the residuals in the velocity field, Fig.~\ref{vel.residuals}). In addition, a systematic uncertainty exists which depends on the inclination of the disk. Following \citet{begemann87}, the systematic uncertainty in inclination for NGC~1167 is 2\degr , which is constant with radius ($i=38$\degr $\pm 2$\degr). This translates into a possible systematic offset of 15~\kms . The position angle is constant within the uncertainties ($PA=75$\degr $\pm 3$\degr ). The remarkable bump seen in Fig.~\ref{pv.slices} is also visible in the rotation curve as a distinct feature at radii between 80\arcsec\ and 130\arcsec . In principle, the feature in the kinematics could also be explained as the result of a change in inclination instead of rotation velocity. However, a bending up and down in inclination of $\sim$6\degr\ would be required. Such a bending is not observed in other galaxies and therefore we think that it is unlikely that this feature reflects such a change in inclination.

%
%
%
%
%
\begin{table*}
\caption{Parameters for mass models with isothermal halos, \HI scaling and MOND.}
\label{tab.massmodels}
\centering
\begin{tabular}{l c c c r c c c}
\hline\hline
model & $\Gamma_b$ & $\Gamma_d$ & $\eta$ & $r_c$ & $\rho_0$ & $A$ & $\chi^2$\\
 & \Msun/\Lsun & \Msun/\Lsun &  & kpc & \Msun pc$^{-3}$ &  & \\
(1) & (2) & (3) & (4) & (5) & (6) & (7) & (8)\\
\hline
minimal $\chi^2$  & 4.1 & 3.6 &  1.4 & 10.6 &  14.5 &    - & 12.2\\
max. allowed disk & 4.3 & 4.7 &  1.4 & 25.6 &   3.0 &    - & 16.3\\
min. allowed disk & 3.8 & 1.0 &  1.4 &  2.7 & 249.0 &    - & 18.7\\
constant M/L      & 4.2 & 4.2 &  1.4 & 21.0 &   4.4 &    - & 16.6\\
\HI scaling       & 4.5 & 4.9 & 46.4 &    - &     - &    - & 11.2\\
MOND              & 3.8 & 4.9 &  1.4 &    - &     - & 3565 & 13.8\\
\hline
\end{tabular}
\caption*{Note. -- (1) model; (2) bulge $R$-band mass-to-light ratio; (3) disk $R$-band mass-to-light ratio; (4) \HI scaling factor; (5) halo core radius; (6) halo central density; (7) Mond factor; (8) $\chi^2$ value from fit.}
\end{table*}
%

\subsection{Large-scale streaming motions}

The presence of non-circular motions is also evident from the residual velocity field. In
Figure~\ref{vel.residuals} (top panel) the velocity field of an axisymmetric model, which is based on our derived rotation curve, has been subtracted from the observed velocity field. A large-scale residual structure is seen at the radius of the feature in the rotation curve. We note that the residuals are not completely axisymmetric, in agreement with the slightly asymmetric appearance of the feature in the rotation curve (Sect.~3.1). This residual structure is even more prominent assuming a purely flat model rotation curve (i.e. without a bump), as shown in the bottom panel of Fig.~\ref{vel.residuals}. Additional velocity deviations from the axisymmetric model at larger and smaller radii are also present.

The systematic residuals imply that streaming motions must be present in the disk. Simple radial motions are excluded because the feature in the rotation curve is seen strongest along the major axis (where radial motions should be zero) and not along the minor axis (where they should be maximal). The measured peak velocity of the large-scale residuals is above 34~\kms\ corresponding to a deprojected amplitude (i.e. velocity in the plane of the disk) of $55$~\kms . In Sect.~6.3 we argue that these large-scale streaming motions may be caused by an interaction/merging event.

%

\subsection{Mass models}

We model the mass distribution in order to investigate whether the feature in the rotation curve can be explained by one of the standard models. We assume that the gravitational potential of a disk galaxy can be written as the sum of the individual components: stellar bulge, stellar disk, gaseous disk (multiplied with the \HI scaling factor $\eta$) and a dark matter halo. To model the stellar mass distribution we use published $R$-band photometry \citep{noordermeer06} and adopt Noordermeer's disk/bulge separation. The uncertainty of the rotation curve affects the $M/L$ estimate by $\le 0.3$.

\begin{figure*}
\centering
\includegraphics[width=0.33\textwidth, angle=270]{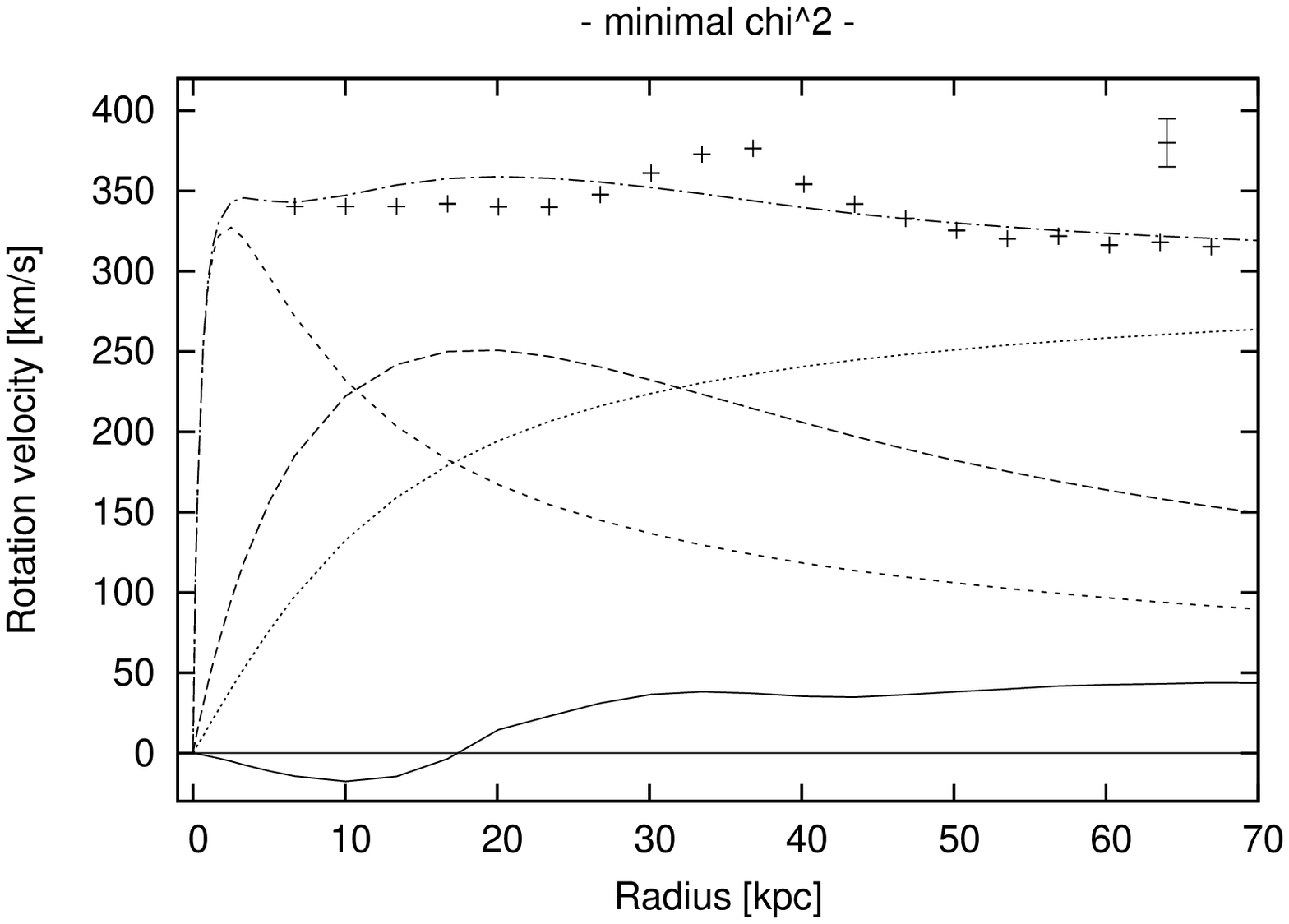}
\includegraphics[width=0.33\textwidth, angle=270]{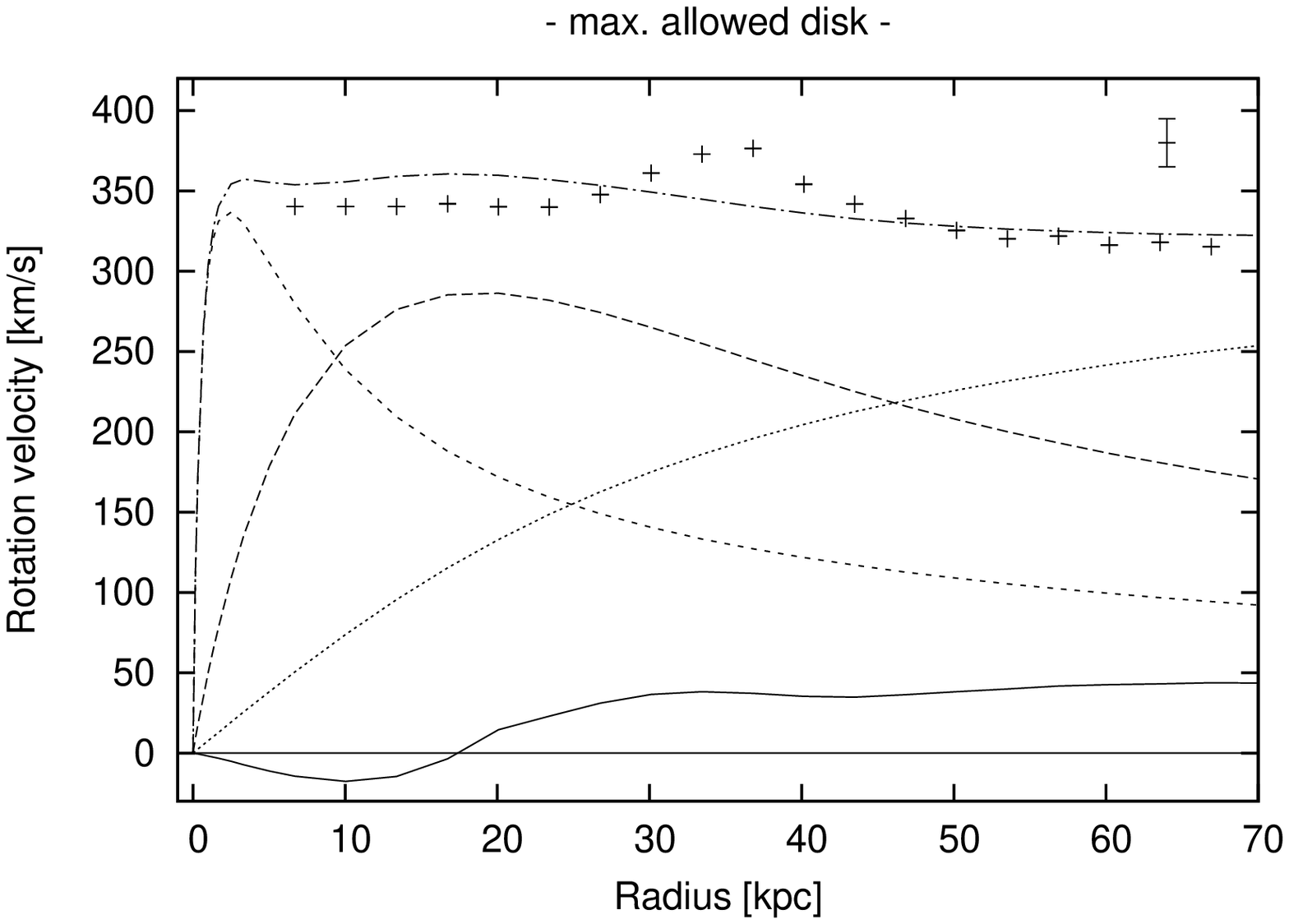}
\includegraphics[width=0.33\textwidth, angle=270]{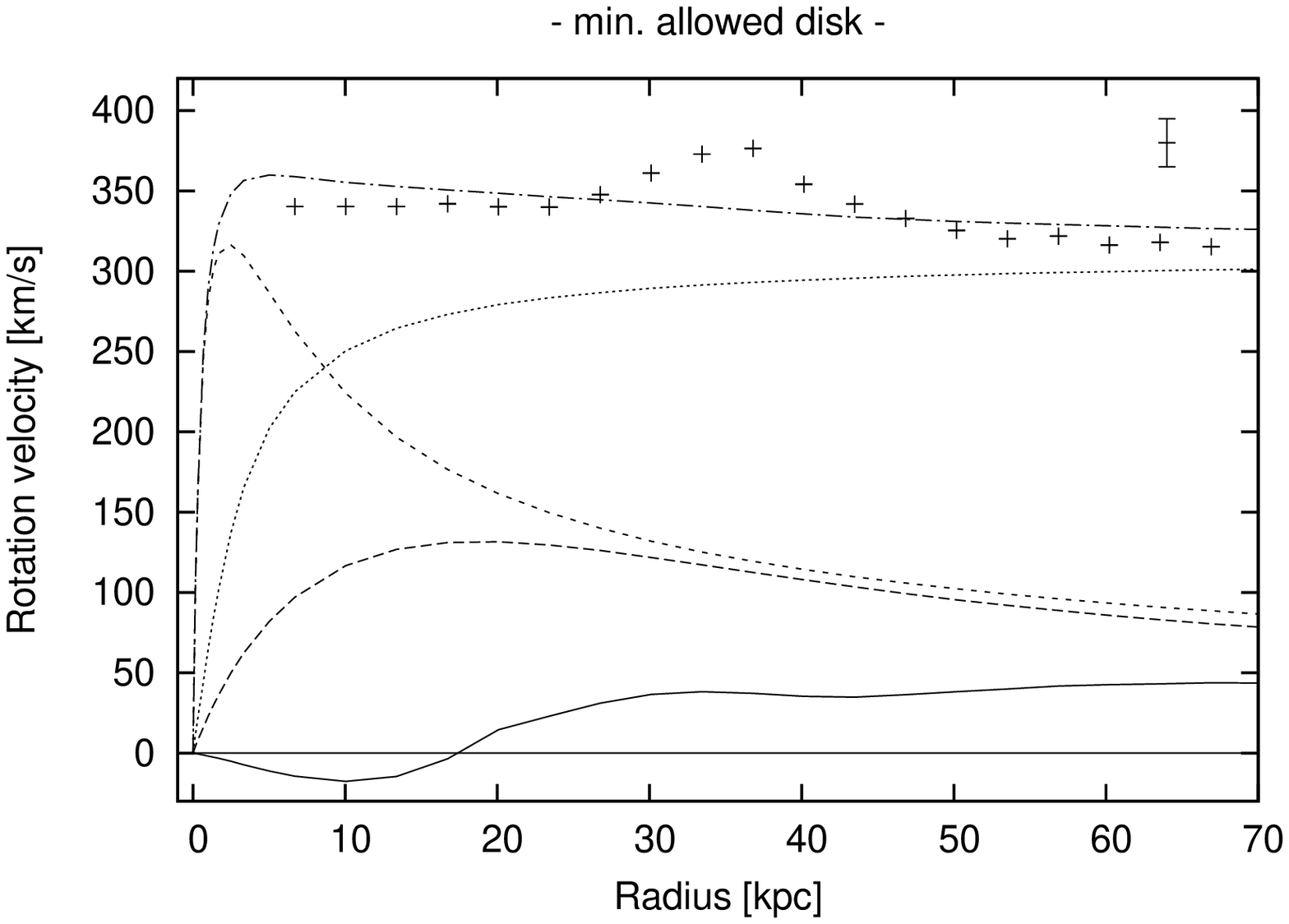}
\includegraphics[width=0.33\textwidth, angle=270]{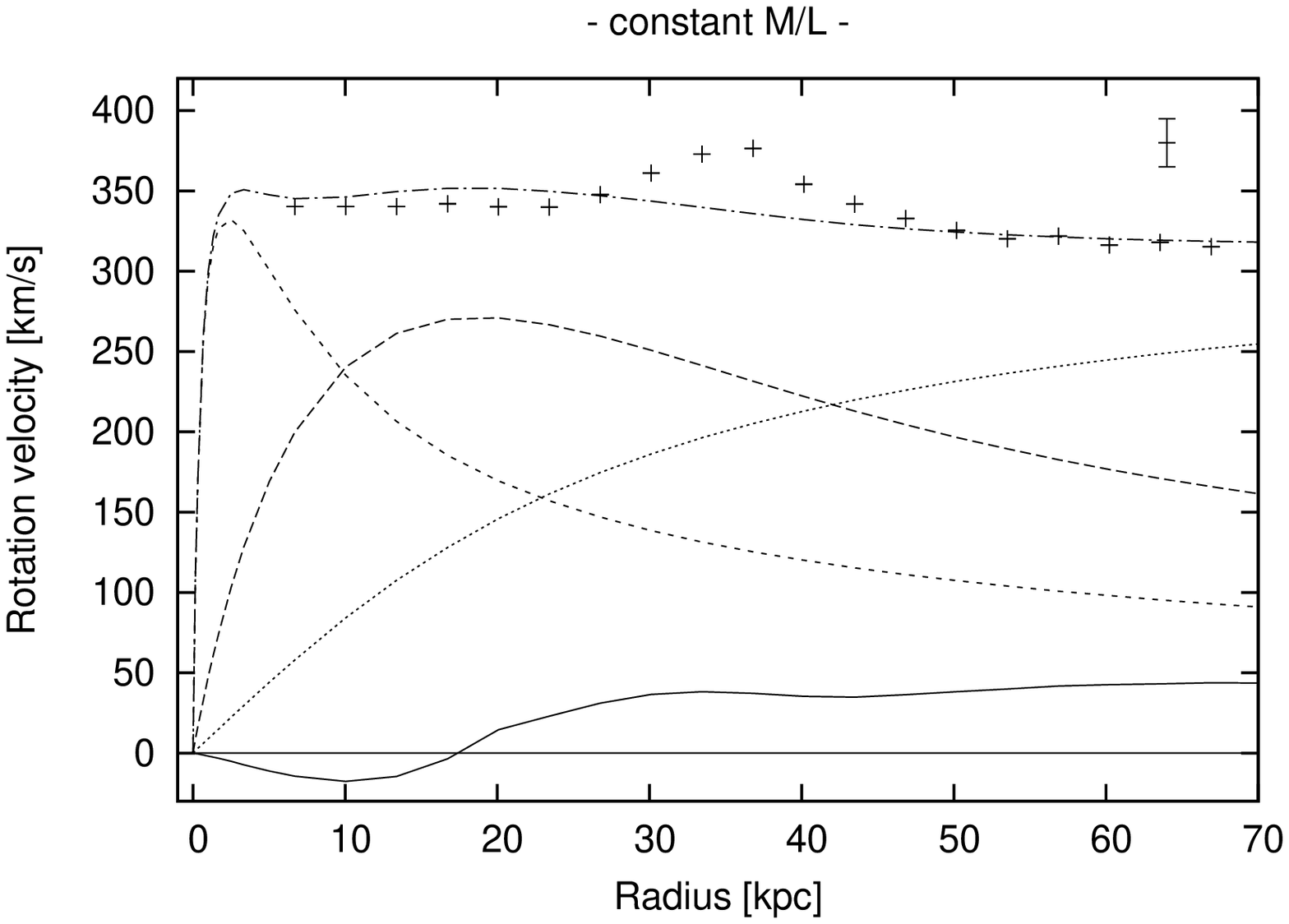}
\includegraphics[width=0.33\textwidth, angle=270]{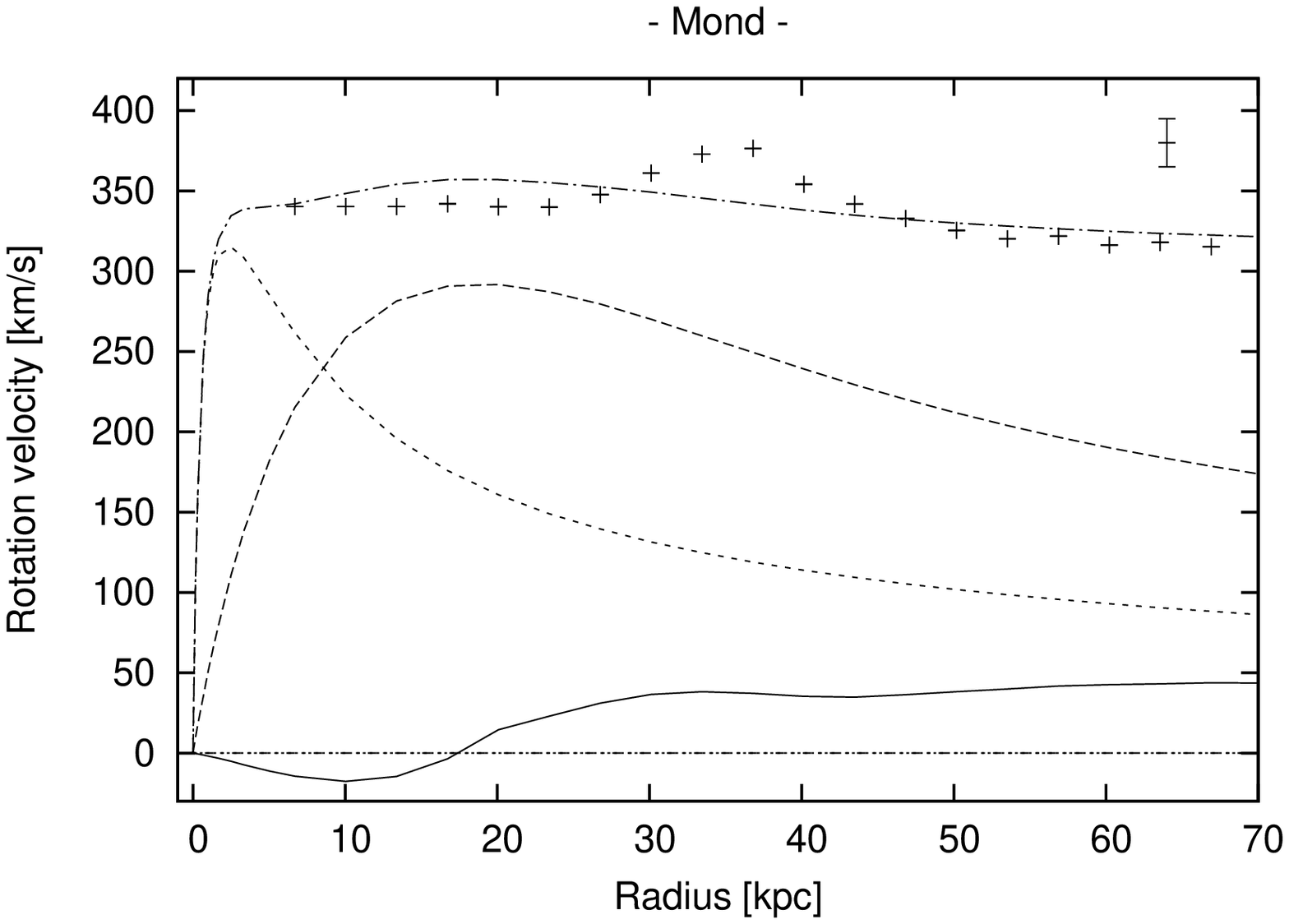}
\includegraphics[width=0.33\textwidth, angle=270]{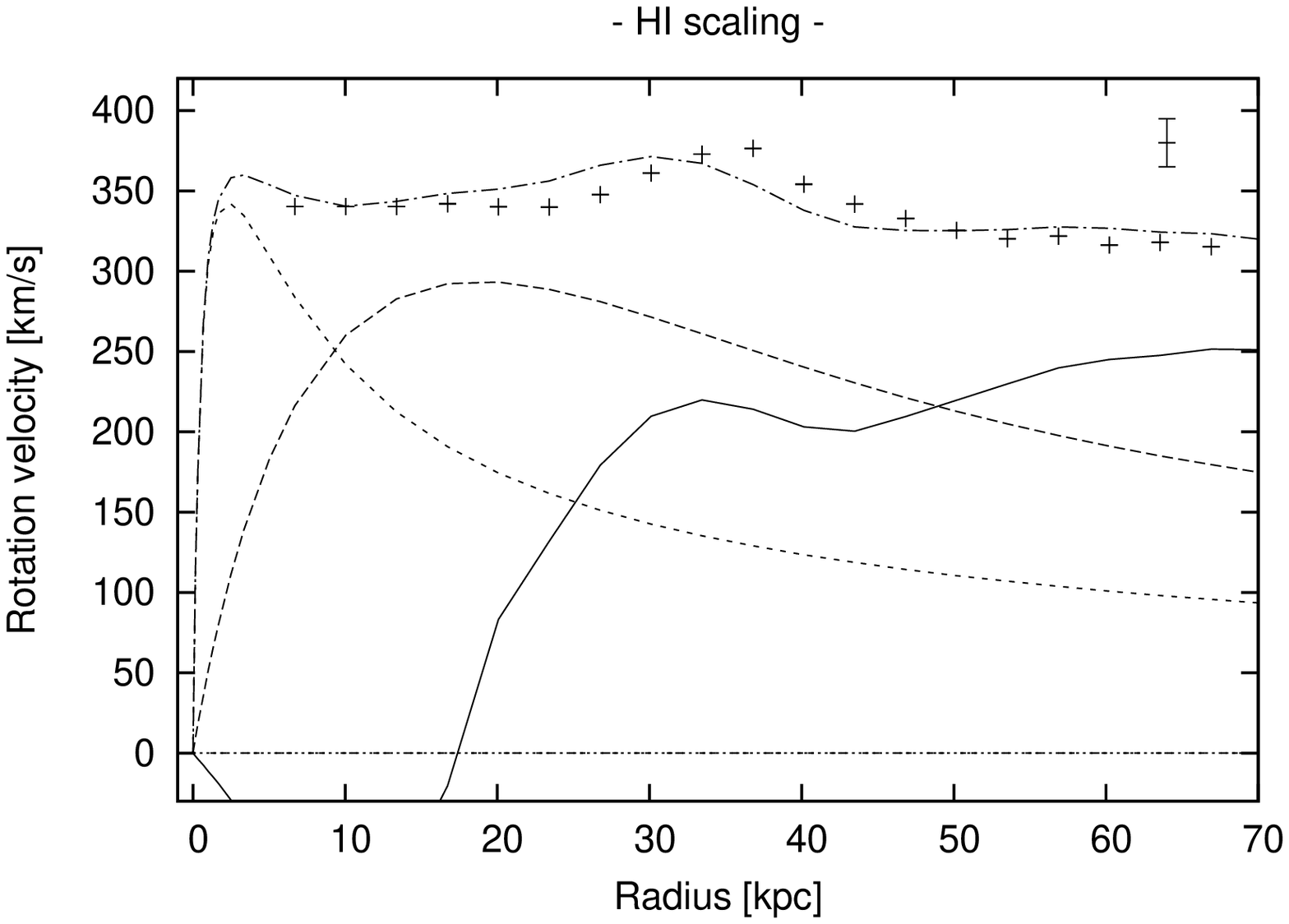}
\caption{Mass models. The fitted individual contributions to the rotation curve are stellar bulge (short dashed), stellar disk (long dashed), gas disk (continuous line) and dark halo (dotted). The dot-dashed line shows the resulting model rotation curve. The error bar indicates the uncertainty for an individual data point of the observed rotation curve.}
\label{mass.models}
\end{figure*}

None of the tested models reproduces the feature in the rotation curve satisfactorily (see Fig.~\ref{mass.models} and Tab.~\ref{tab.massmodels}). A truncated optical disk \citep[see, e.g.,][]{vanderkruit07} is not observed and, therefore, neither explains the feature in the rotation curve. Moreover, even if a truncation in the optical disk would have been observed at the right radius, it would not have been able, given the faintness of the disk, to explain the amplitude of the bump.

The best model (smallest overall $\chi^2$) is that of a scaled-up version of the \HI mass distribution with no dark matter halo. The \HI scaling factor of 33 (correcting for helium) is higher than the average value that \citet{hoekstra01} found for a sample of 24 spiral galaxies ($\eta =6.5$), but is in agreement with the systematic trend that more luminous systems require higher scaling factors than low luminosity systems. High scaling factors have also been found for other luminous galaxies, e.g., $\eta=37$ for UGC~6787 \citep{noordermeer06}. However, also in the \HI scaling model the feature in the rotation curve is not perfectly reproduced. The predicted bump is approximately $5$~kpc closer to the centre than the observed bump.

The fact that none of the tested models fits the observed rotation curve suggests that the feature in the rotation curve does not reflect an equilibrium mass distribution of NGC~1167. In Sect.~6.3 we will argue that the bump in velocity is consistent with the presence of large-scale streaming motions in the disk, as already suggested from the residual velocity field (Fig.~\ref{vel.residuals} and Sect.~5.1).

%
%

\section{Discussion}

We start this section by discussing the formation of the \HI disk and the role of different accretion mechanisms (Sect.~6.1). In Sect.~6.2 we give possible explanations for the absence of a population of cold gas clouds and Sect.~6.3 briefly comments on a possible connection between the \HI density and the large-scale streaming motions in the disk.

\subsection{Formation of the \HI disk}

Where does the \HI in NGC~1167 come from? The various signs of merging and interaction activities suggest that currently the \HI is brought in by fairly massive satellite galaxies. This means that most of the \HI arrives in large quantities in a few events.

Signs of recent and/or ongoing interactions with the environment of NGC~1167 are, for instance, the southwestern extension of the \HI disk, the feature in the rotation curve, the disturbed \HI morphology and kinematics of UGC~2465 and the disrupted satellite at the northern edge of the \HI disk (Sect.~3.2). All of these signs are pointing towards an ongoing assembly of the \HI disk. About 30\% of the detected \HI in the data cube is located in companions. Hence, it might be possible that, over the course of several Gyr, NGC~1167 has built-up a large fraction of its \HI disk via minor mergers and interactions with the environment.

The southwestern extension could be the result of such a minor merger. Figure~\ref{deep.optical} shows the disrupted satellite at the northern edge of the \HI disk with a possible tail-like structure all the way to the \HI extension (Sect.~3.2). Therefore, we might see also the disrupted \HI in the aftermath of a minor merger. The \HI mass of the southwestern extension is at least $9\times 10^8$~\Msun , showing that the accretion of satellite galaxies can bring in large amounts of \Hi . This minor merger may have also induced the large-scale streaming motions in the disk that are seen as the feature in the rotation curve (Sect.~5.1, see also Sect.~6.3). The non-circular motions provide, therefore, another piece of evidence for the accretion of, fairly massive, companion galaxies.

Another possibility is that the \HI extension could be caused by a recent interaction with UGC~2465. In this case, the past interaction time estimate is roughly 1~Gyr, given the projected distance between both galaxies and assuming a relative velocity offset of 200~\kms . This time estimate equals one revolution of NGC~1167 (at $r=65$~kpc), but is less than the 1.3~Gyr that one revolution of UGC~2465 takes. After an interaction it takes a few rotations before the gas settles in a rotating disk structure. Hence, it is not a surprise that the southwestern extension and part of the \HI structure of UGC~2465 is not completely settled.


Other mechanisms that could contribute to the \HI disk building process do not seem to play a major role. Below, we briefly discuss the relevance of small gas clouds, the possibility of a major merger and the role of stellar mass loss.

\vspace{0.3cm}

\noindent {\sl Accretion of small gas clumps:}\\
We do not find a large number of small gas clumps. In fact, only two clouds just outside the \HI disk  (No. 4 and 5 in Tab.~\ref{neighbours}) are detected. The very few tentative \HI structures (with masses of $\sim$$10^7$~\Msun ), which are not explained by the model (Sect.~4), are features at the noise level and form an upper limit for a possible halo cloud population. The two detected clouds have masses at the high mass end of the high velocity clouds (HVCs) observed around the Milky Way and other nearby galaxies \citep[$10^4 \le M_{\rm{HI}} \le $$10^7$~\Msun; e.g.][]{thilker04, westmeier05, wakker07, wakker08}. Clouds with \HI masses $\sim$$10^7$~\Msun\ were also detected around NGC~2403 \citep{fraternali02}, NGC~891 \citep{oosterloo07}, NGC~6946 \citep{boomsma08} and NGC~2997 \citep{hess09}. What makes these HVCs interesting is that they could be, at least in some cases, gas condensations from the IGM. The presence of two such clouds around NGC~1167 might, therefore, indicate that the accretion of small gas clouds, at a low level, occurs.

Nevertheless, the absence of a large number of such clouds shows that the ``diffuse'' accretion of small cold gas clumps from the IGM is not the dominant disk building process. The number of required clumps ($>10^3$) would be unphysically large, given that we detect only very few such gas clouds. Possible explanations for the absence of gas clouds located in the halo are discussed in Sect.~6.2.

\vspace{0.3cm}

\noindent {\sl Major merger:}\\
While gas-rich major mergers can form giant low column density \HI disks (Sect.~1), we have no evidence in our \HI data that this is actually the case for NGC~1167. Neither do the deep optical images show any indication of a major merger. Such a dramatic event would certainly need to be older than a few Gyr, given that the \HI disk shows regular rotation for $r<65$~kpc. After a major merger, it takes at least a few revolutions (corresponding to a few Gyr at $r=65$~kpc) before the gas settles in a regular disk structure. A stellar population analysis of \citet{emonts06} also suggests that the last interaction (that caused a burst of star formation ) occured at least one to several Gyr ago, but this starburst may be the result of a smaller accretion event and is no evidence for a major merger.

\vspace{0.3cm}

\noindent {\sl Stellar mass loss:}\\
The \HI can not originate from stellar mass loss. All stars shed mass over their lifetimes. According to models of \citet{ciotti91} and \citet{brighenti00}, the expected cold gas mass from stellar mass loss processes for massive galaxies could be $>10^{10}$~\Msun\ over a Hubble time. In the case of NGC~1167 it is {\sl unlikely} that the \HI results from stellar mass loss because the \HI disk extends well beyond the optical disk ($D_{\rm{HI}}=3\times R_{25}$) and carries therefore more angular momentum than the stars. Also in other galaxies there is little evidence that cold gas comes from mass loss processes. In the cases where gas could be associated with stellar mass loss, it is ionised \citep[see, e.g.,][]{sarzi06}.

\vspace{0.3cm}



\noindent As a final remark, we would like to note that all eight detected companions of NGC~1167 have systemic velocities close to that of NGC~1167 (none is deviating by more than 200~\kms, see Tab.~\ref{neighbours}). Furthermore, all but No.~7 (in Tab.~\ref{neighbours}) have systemic velocities which follow the sense of rotation of NGC~1167. We cannot rule out that this is the result of projection effects. However, \citet{erickson99} found the trend that companions seem to be close in velocity to the host galaxy in a number of (late-type) spiral galaxies. They argue that their sample is not affected by projection effects and they find evidence for varying sizes of dark matter halos. However, for galaxy sizes above 200~kpc their constraints become weak. Larger statistical samples are needed to shed more light on this issue.

\subsection{The absence of an \HI halo}

Different theoretical explanations have been proposed that could explain the absence of an extended halo of cold gas clouds. In principal, small cold gas clouds from the IGM can be accreted onto the disks of galaxies \citep[][see also Sect.~1]{binney77}. However, the surrounding x-ray halo of massive galaxies ionises the cold gas before it can reach the disk \citep[e.g.][]{nipoti07}. According to simulations of e.g. \citet{dekel08}, accretion of cold gas is not expected at galaxy masses above $\sim 10^{12}$~\Msun. The dynamical mass of NGC~1167 is at least $M_{\rm{dyn}} > 1.6 \times 10^{12}$~\Msun\ (for $r_{\rm{max}}=65$~kpc and $v_{\rm{rot}}=330$~\kms ). Hence, NGC~1167 may simply be too massive for the accretion of small gas clumps.

Another possible explanation for the absence of an extended halo of cold gas clouds in NGC~1167 is the hypothesis that cold halos are caused, directly and indirectly, by star formation.  Because NGC~1167 does not form stars, galactic fountains from supernovae explosions do not blow any gas into the halo. However, according to e.g. \citet{marinacci10}, this galactic fountain gas would be needed to create Kelvin-Helmholtz instabilities in the hot corona such that (hot) gas can condense to clouds that are able to reach the disk. With the presence of supernovae, star-forming galaxies can accumulate enough cold gas to sustain their observed star formation rates \citep[see also][]{hopkins08} while in early-type galaxies, like NGC~1167, the creation of instabilities in the halo is prevented (due to the absence of galactic fountains). Hence, no new cold gas complexes can reach the disk of NGC~1167.

\subsection{Streaming motions, rotation curve and \HI density distribution}

Streaming motions in disks are a well known phenomenon and observed in many galaxies \citep[see e.g.][]{knapen97,pisano98,fathi06}. However, it is remarkable that the streaming motions in NGC~1167 are (to first order) symmetric. The amplitude is relatively large ($\approx 50$~\kms , deprojected), at the high end of what is detected in a number of other early-type galaxies \citep{noordermeer06}. Interestingly, most of these galaxies, like NGC~1167, also show clear signs of interaction and/or tidal distortions with some of them having bumps and wiggles in the rotation curve as well \citep[see also][]{noordermeer07}.

The feature in the rotation curve of NGC~1167 occurs at the radius where the \HI surface density drops significantly (see Fig.~\ref{density} and \ref{pv.slices}). The drop in surface density results in the bump in the \HI scaling model rotation curve (Fig.~\ref{mass.models} and Sect.~5.1), suggesting that a connection between large-scale streaming motions and \HI density distribution exists.

\HI scaling has previously been tried in the analysis of rotation curves and has even led to suggestions of a possible coupling of dark matter with the gas component \citep[see, e.g.,][]{broeils92, swaters99, hoekstra01, noordermeer06}. Remarkably, in galaxies with bumps and wiggles, the rotation curves are fitted best by models with scaled \HI disks \citep[e.g. in UGC~2916, UGC~5253, UGC~6787 and UGC~11852; see][]{noordermeer06}. We therefore suspect that in many galaxies a connection between \HI density distributions and streaming motions in the disks may exist -- because both are the result of an interaction. This may be the reason for the apparent success of \HI scaling in describing such rotation curves.


\section{Summary}

We reported deep WSRT \HI observations of the early-type galaxy NGC~1167. Our main conclusions are:

(i) NGC~1167 is a very massive early-type galaxy with a giant \HI disk ($D_{\rm{HI}} =160$~kpc). The \HI mass of $1.5 \times 10^{10}$~\Msun\ contributes about 1\% to the total mass, the \HI surface density ($\le 2$~\Msun ~pc$^{-2}$) is too low for starformation to occur.

(ii) The \HI disk shows regular rotation for $r<65$~kpc with a flat rotation curve of $\sim$340~\kms\ amplitude. At $r\approx 1.3\times R_{25}$ the rotation curve shows a bump of $\sim$50~\kms\ on the approaching and receding side which is due to large-scale streaming motions in the disk. We argue that both the streaming motions and density features are due to interactions and that this explains the success of \HI scaling in fitting rotation curves.

(iii) NGC~1167 shows various signs of recent satellite accretion, interaction and merging activities as well as eight \HI companions (with $r<350$~kpc). No extended halo of cold gas clouds is detected with column densities above $6\times 10^{18}$~cm$^{-2}$. This suggests that the \HI disk is currently primarily formed through the accretion of companion galaxies which bring in large amounts of \HI to NGC~1167 and that the accretion of small cold gas clumps is not an important disk building mechanism in this massive galaxy.

In a future paper (Struve et al. in prep.) we will discuss whether our results, for instance the presence of streaming motions in the disk, also have implications for the current phase of nuclear activity of the radio source B2~0258+35. The origin and onset of nuclear activity in galaxies is still not fully understood. Gas certainly plays an important role in the triggering of the black hole and NGC~1167 is a suitable object to study the interplay between the host galaxy and the radio source.


\begin{acknowledgements}

We thank the referee Jacqueline van Gorkom for her helpful suggestions which improved the paper.
CS was supported during this research by the EU Framework 6 Marie Curie Early Stage Training programme under contract number MEST-CT-2005-19669 ESTRELA.
The WSRT is operated by the Netherlands Institute for Radio Astronomy with support from the Netherlands Foundation for Scientific Research. 
This research has made use of the NASA/IPAC Extragalactic Database.
The Digitized Sky Survey was produced at the Space Telescope Science Institute under US government grant NAG W-2166.

\end{acknowledgements}

\Online
\begin{appendix} 
\section{Channel maps}

%
\begin{figure*}[h]
\centering
\includegraphics[width=0.7\textwidth]{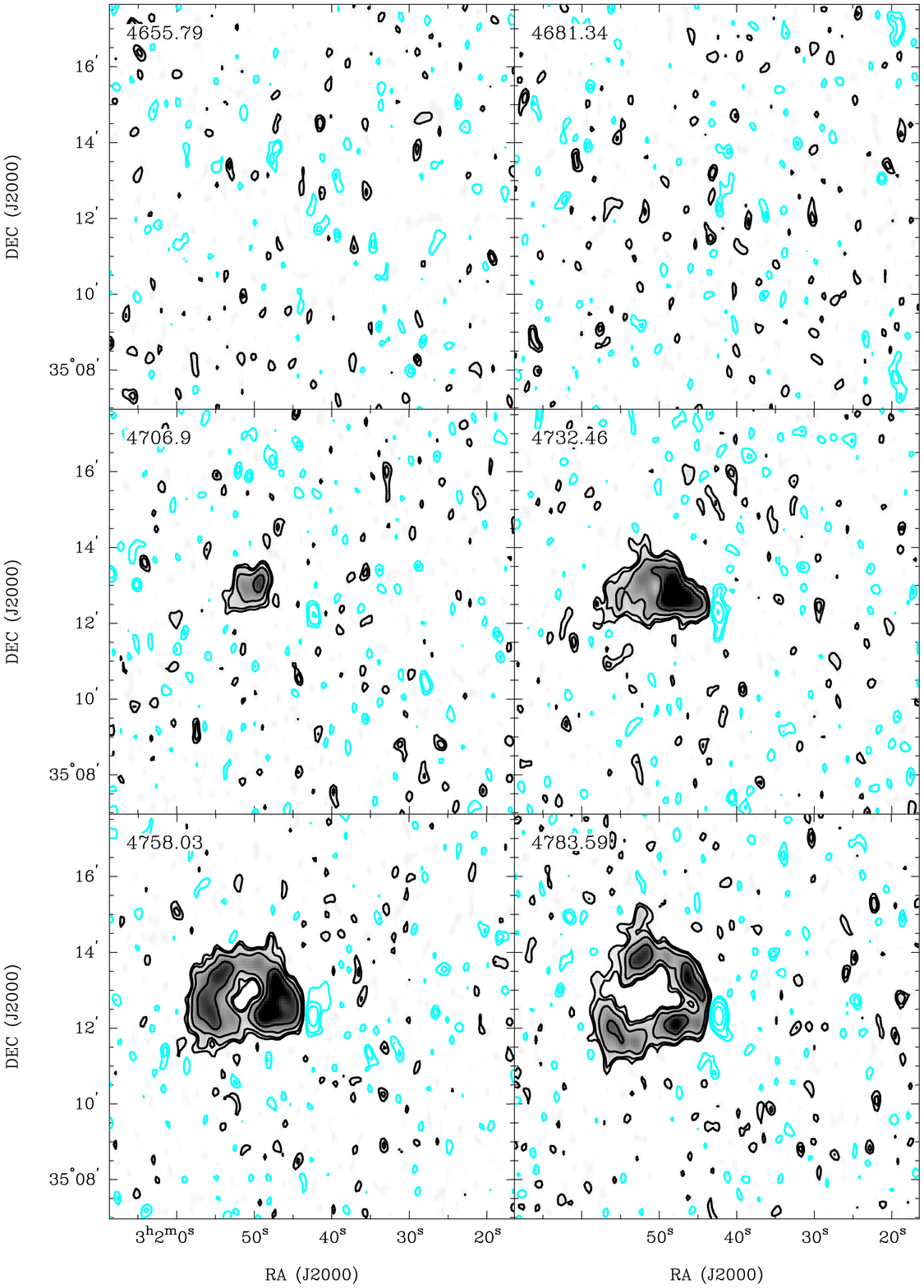}
  \vspace{35pt}
\caption{Channel maps showing every third velocity channel of the data cube. Contour levels: -24, -12, -6, -3, -2 (turquoise), 2, 3, 6, 12 and $24\sigma$ (black).}
\label{channelmaps}
\end{figure*}
%
%
\begin{figure*}
  \ContinuedFloat
\centering
\includegraphics[width=0.7\textwidth]{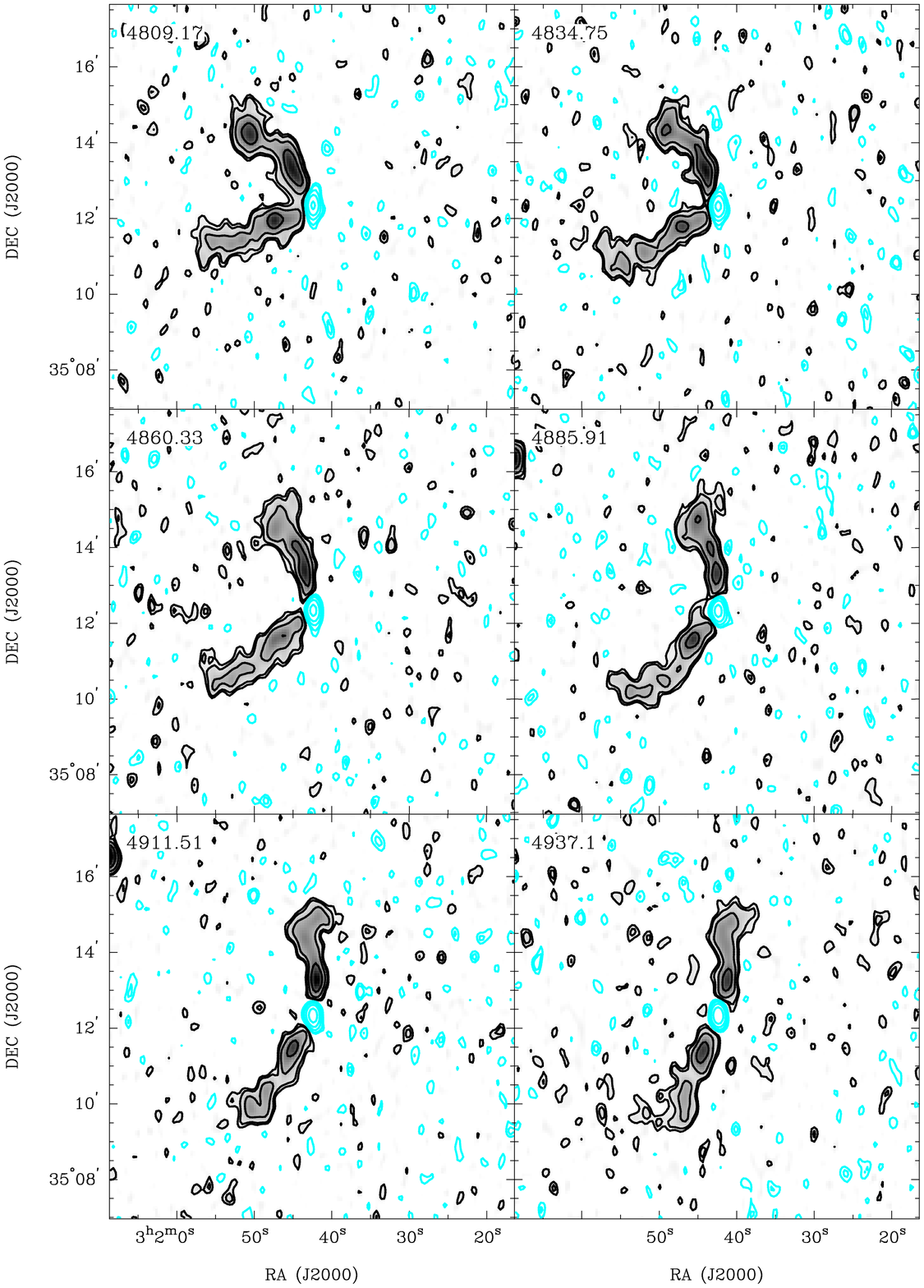}
  \vspace{35pt}
\caption{--- {\sl Continued.}}
\label{channelmaps}
\end{figure*}
%
%
\begin{figure*}
  \ContinuedFloat
\centering
\includegraphics[width=0.7\textwidth]{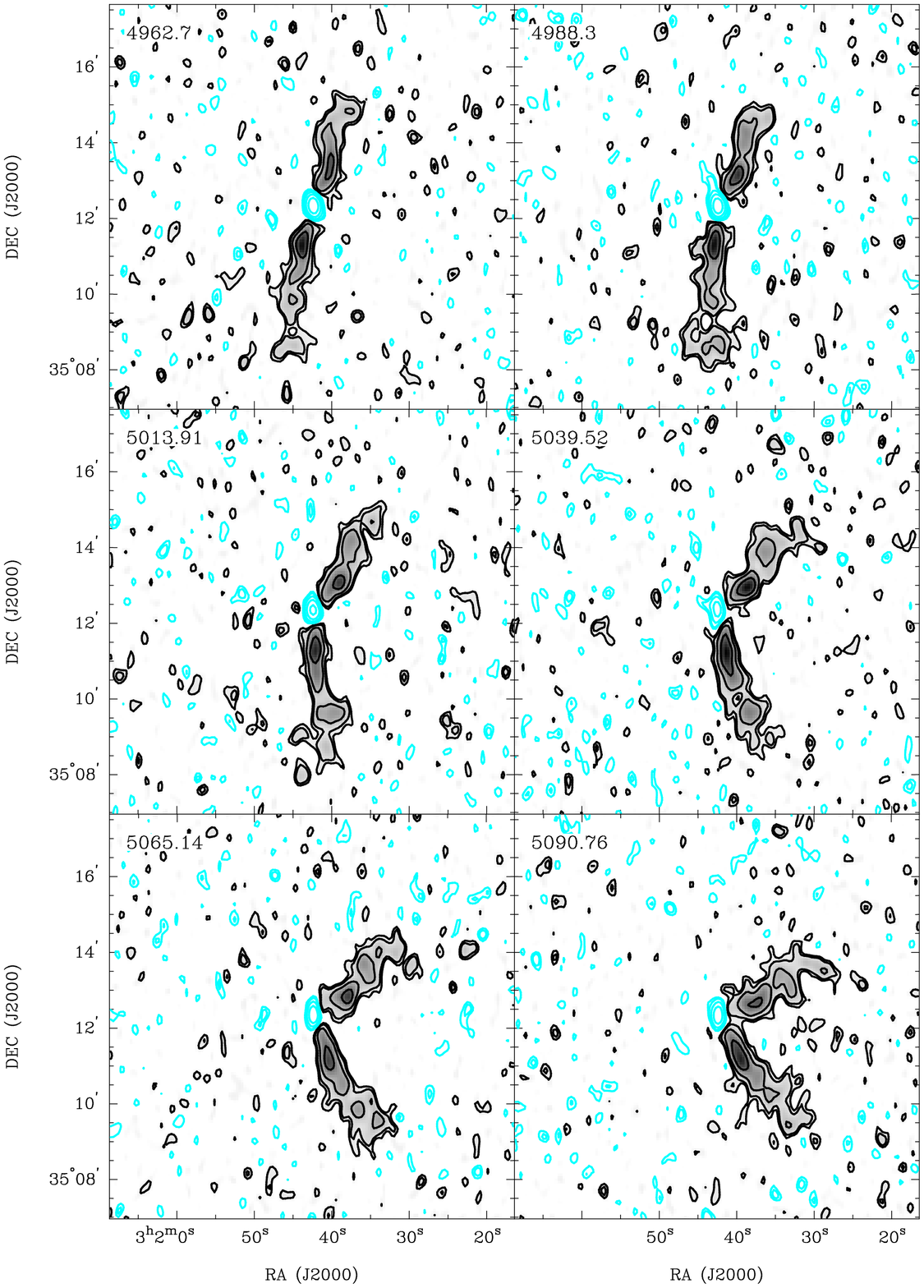}
  \vspace{35pt}
\caption{--- {\sl Continued.}}
\label{channelmaps}
\end{figure*}
%
%
\begin{figure*}
  \ContinuedFloat
\centering
\includegraphics[width=0.7\textwidth]{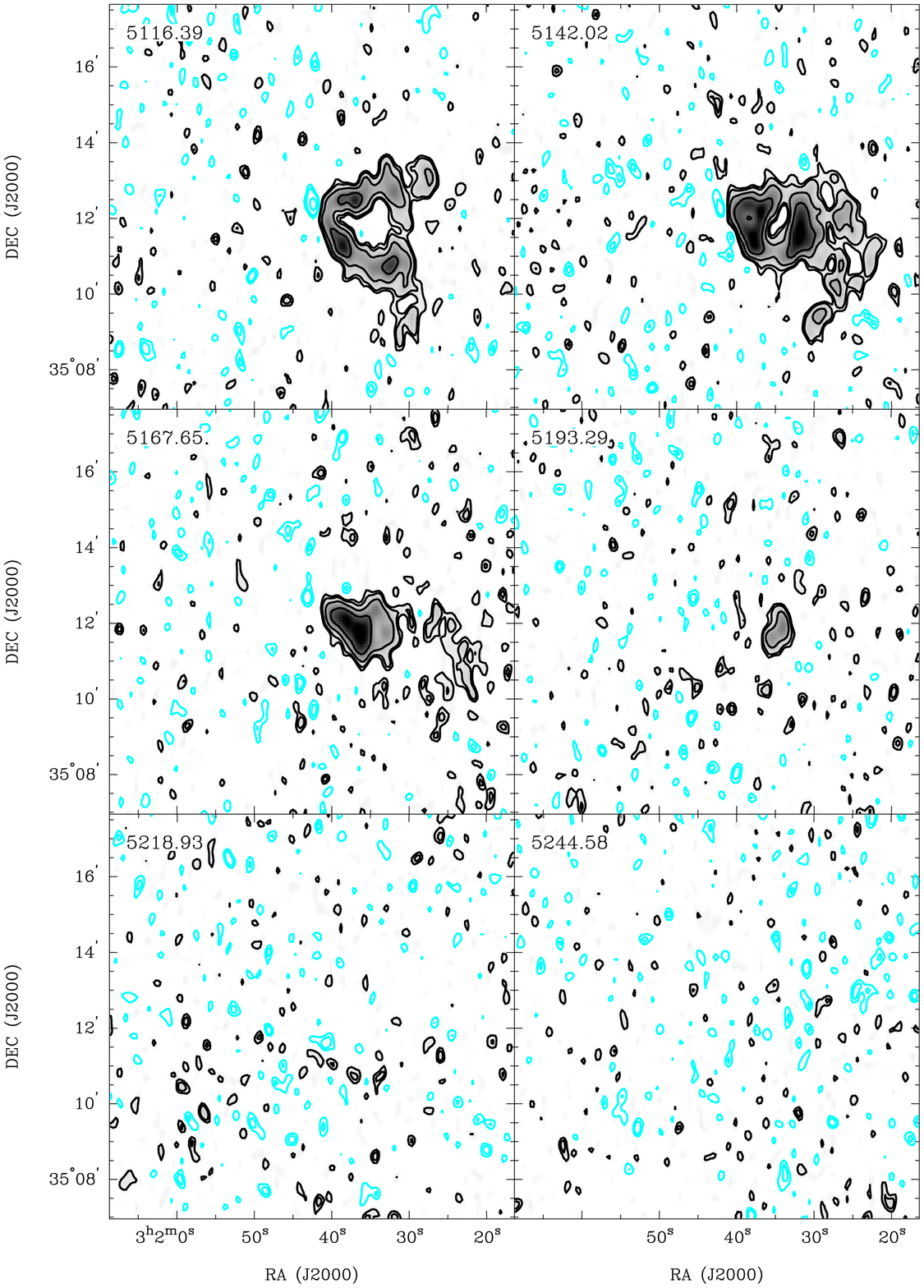}
  \vspace{35pt}
\caption{--- {\sl Continued.}}
\label{channelmaps}
\end{figure*}
%

\newpage

\section{Table: Rotation curve}

%
\begin{table}
\caption{Rotation curve.}
\label{RC}
\centering
\begin{tabular}{ c c c}
\hline\hline
Radius [\arcsec] & Radius [kpc] & velocity [\kms ]\\
\hline
  20 &   6.7 & 334   \\ 
  30 & 10.0 & 335   \\
  40 & 13.3 & 335   \\
  50 & 16.7 & 337   \\
  60 & 20.0 & 340   \\
  70 & 23.3 & 340   \\
  80 & 26.7 & 348   \\
  90 & 30.0 & 361   \\
100 & 33.3 & 373   \\
110 & 36.7 & 377   \\
120 & 40.0 & 354   \\
130 & 43.3 & 342   \\
140 & 46.7 & 333   \\
150 & 50.0 & 325   \\
160 & 53.3 & 320   \\
170 & 56.7 & 322   \\
180 & 60.0 & 316   \\
190 & 63.3 & 318   \\
200 & 66.7 & 315   \\
\end{tabular}
\end{table}
%

\end{appendix}

\end{document}